\documentclass[11pt]{article}
\usepackage{amssymb,amsmath,mathrsfs}
\usepackage{graphicx}

\newtheorem{theorem}{Theorem}[section]

\begin{document}


\title{Multi-soliton, multi-positon, multi-negaton, and multi-periodic solutions of
the coupled Volterra lattice equation }

\author{Hai-qiong Zhao \qquad Zuo-nong Zhu\\
Department of Mathematics, Shanghai Jiao Tong University,\\
800 Dongchuan Road, Shanghai, 200240, P.R. China}
\date{\em 15 November  2009}
\maketitle

\begin{abstract}
This paper aims to find new explicit solutions including
multi-soliton, multi-positon, multi-negaton, and multi-periodic
for a coupled Volterra lattice system which is an integrable
discrete version of the
coupled KdV equation. The dynamical properties of these new solutions are discussed in detail.
\end{abstract}




\section{Introduction}
~~~~~~ As is well known, in recent years there has been an
explosion of interest in the study of discrete integrable systems.
This is due to the important role they play in mathematics and
physics and to their many applications. The Volterra equation
\begin{equation}\label{volterra}
\frac{d U_n}{dt}+U_n(U_{n+1}-U_{n-1})=0
\end{equation}
is an important lattice. It has been shown that the Volterra
equation possesses all useful integrable properties. For example,
it can be solved by the inverse scattering transform
\cite{volterra_inver}; it has the Darboux transformation and
various solutions \cite{volterra_solution}; it possesses
infinitely many generalized symmetries and infinitely many
integrals of motion \cite{toda_sym1}-\cite{toda_sym3}; it gives
the KdV equation in the continuous limit. As we known, the topic
on the relations between discrete integrable systems and KdV-type
theories has attracted much researches \cite{Sch}-\cite{conti2}.
Here we quote Morosi and Pizzocchero's monograph \cite{conti1}. In
their paper, KdV theory including the infinitely many commuting
vector fields, the conserved functions, the Lax pairs and the
bi-Hamiltonian structure is recovered systematically through the
continuous limit of the Kac-Moerbeke (KM) system. Aiming to get
more insight on the relation between Volterra-type lattice and
KdV-type equation, recently Lou et al \cite{Cvolterra} introduced
a coupled Volterra system
\begin{eqnarray}\label{Cvolterra}
 \begin{array}{ll}
 \dot{a}_n+a_n(a_{n+1}-a_{n-1})-\alpha b_n(b_{n-1}-b_{n+1})=0, \\
 \dot{b}_n+b_n(a_{n+1}-a_{n-1})+a_n(b_{n+1}-b_{n-1})=0,
  \end{array}
\end{eqnarray}
which yields the two-coupled KdV equation,
\begin{equation}\label{CKdV}
 \begin{array}{ll}
 u_t+6\alpha vv_x+6uu_x+u_{xxx}=0, \\
 v_t+6vu_x+6uv_x+v_{xxx}=0,
  \end{array}
\end{equation}
through the continuous limit
\begin{eqnarray}
a_n=1+\delta^2 u((n-2t)\delta,\frac{1}{3}\delta^3t),\qquad
b_n=\delta^2 v ((n-2t)\delta,\frac{1}{3}\delta^3t).
\end{eqnarray}
Equation \eqref{CKdV} has some physical applications including the
atmospheric dynamics, Bose-Einstein condensation and two-wave
modes in a shallow stratified liquid \cite{C1}-\cite{C3}. As we
known, finding explicit exact solutions for the discrete
integrable systems is an important and difficult problem. In
Ref.\cite{Cvolterra}, by using a simple function expansion method,
various explicit solutions to equation \eqref{Cvolterra} such as
solitary wave, positon, and complexiton have been given. The
generalized symmetries, recursion operator, and integrability of
coupled Volterra system \eqref{Cvolterra} with $\alpha=-1$ were
given \cite{Cvolterra_symm}.

We first remark here that under transformations
$\sqrt{-\alpha}b_n\rightarrow b_n$ (for the $\alpha<0$ case), or
$\sqrt{\alpha}b_n\rightarrow ib_n$ (for the $\alpha>0$ case),
equation \eqref{Cvolterra} changes to
\begin{eqnarray}\label{Cvolterra2}
 \begin{array}{ll}
 \dot{a}_n+a_n(a_{n+1}-a_{n-1})+b_n(b_{n-1}-b_{n+1})=0, \\
 \dot{b}_n+b_n(a_{n+1}-a_{n-1})+a_n(b_{n+1}-b_{n-1})=0.
  \end{array}
\end{eqnarray}
Although various explicit solutions for equation\eqref{Cvolterra2}
have been given, the more general multi-soliton, multi-positon,
multi-negaton, and multi-periodic solutions have not been
proposed. We think that these explicit solutions can not be given
by using function expansion method. Since coupled Volterra system
(1.2) yields coupled KdV system (1.3) in continuous limit, while
the latter has many physical applications, we think that finding
new explicit solutions to coupled Volterra system
\eqref{Cvolterra2} are important.  This paper is devoted to make
an effort on this topic.  We hope to find new explicit solutions
by using the Darboux transformation.
 It should be remarked that
Darboux transformation not only is a useful method of obtaining
explicit solutions, but also has an important role in mechanics,
physics and differential geometry
\cite{Matveev_d}-\cite{Bobenko_book}. In this paper, we first
construct the Darboux transformation for the coupled Volterra
system \eqref{Cvolterra2} by a special observation. And then, new
explicit solutions for \eqref{Cvolterra2} including multi-soliton,
multi-positon, multi-negaton, and multi-periodic are derived by
using the Darboux transformation. We also analyze some dynamical
properties for these new solutions. \setcounter{equation}{0}
\section{Darboux transformation for the coupled Volterra system \eqref{Cvolterra2}}

~~~~~~In this section, we will construct the Darboux
transformation for the coupled Volterra system \eqref{Cvolterra2}
by a special observation. We can see that the coupled Volterra
system can be read off from the real and imaginary parts of the
complex Volterra system
\begin{equation}\label{mvolterra}
\frac{d u_n}{dt}+u_n(u_{n+1}-u_{n-1})=0,
\end{equation}
where $u_n=a_n+ib_n$. We thus can write the Lax pair of coupled
Volterra system \eqref{Cvolterra2}:
\begin{subequations}\label{Cvolterra_lax}
\begin{equation}
\psi^{(1)}_{n+1}=\lambda \psi^{(1)}_n+b_n \psi^{(2)}_{n-1}-a_n
\psi^{(1)}_{n-1},
\end{equation}
\begin{equation}
\psi^{(2)}_{n+1}=\lambda \psi^{(2)}_n-b_n \psi^{(1)}_{n-1}-a_n
\psi^{(2)}_{n-1},
\end{equation}
\begin{equation}
\frac{d \psi^{(1)}_{n}}{d t}=b_n \psi^{(2)}_n-a_n \psi^{(1)}_{n}+
\lambda a_n \psi^{(1)}_{n-1}- \lambda b_n \psi^{(2)}_{n-1},
\end{equation}
\begin{equation}
\frac{d \psi^{(2)}_{n}}{d t}=-b_n \psi^{(1)}_n-a_n \psi^{(2)}_{n}+
\lambda b_n \psi^{(1)}_{n-1}+ \lambda a_n \psi^{(2)}_{n-1}
\end{equation}
\end{subequations}
from the real and imaginary parts of the complex Lax pair of the
Volterra system
\begin{subequations}\label{volterra_lax}
\begin{equation}
\phi_{n+1}+u_n \phi_{n-1}=\lambda \phi_n,
\end{equation}
\begin{equation}
 \frac{d\phi_n}{dt}=-u_n
\phi_{n}+\lambda u_n \phi_{n-1},
\end{equation}
\end{subequations}
where $ \phi(n,t)=\psi^{(1)}(n,t)+i\psi^{(2)}(n,t)$. Further, we
point out that the $N$-step Darboux transformation of the real
Volterra lattice is applicable to the complex Volterra lattice.

\begin{theorem}\label{Darboux_C}
Equation \eqref{volterra_lax} are invariant with respect to Darboux
transformation
\begin{equation}\label{Darboux_C1}
\phi_n[N]=\frac{W(\phi_{n},\phi_{N,n}, \phi_{N-1,n}, ...,
\phi_{1,n})}{W(\phi_{N,n-1}, \phi_{N-1,n-1}, ..., \phi_{1,n-1})},
\end{equation}
where $\phi_{i,n}$ is a fixed solution of \eqref{volterra_lax}
taken at the point $\lambda=\lambda_i$ $(i=1, 2, 3, ..., N)$, and
$\phi_n$ is a solution of \eqref{volterra_lax} for arbitrary
spectral parameter $\lambda$; where

\begin{equation}\label{Darboux_C2}
\begin{array}{ll} & W(\phi_{N,n-1}, \phi_{N-1,n-1}, ..., \phi_{1,n-1})\\
&=\begin{vmatrix}
 \lambda^{N-1}_N \phi_{N,n-1} &
\lambda^{N-1}_{N-1} \phi_{N-1,n-1}& ... &
\lambda^{N-1}_{1} \phi_{1,n-1}  \\
 \lambda^{N-2}_N \phi_{N,n-2} &
\lambda^{N-2}_{N-1} \phi_{N-1,n-2}& ... &
\lambda^{N-2}_{1} \phi_{1,n-2}  \\
 ... & ... & ... & ...  \\
\lambda_N \phi_{N,n-N+1} & \lambda_{N-1} \phi_{N-1,n-N+1}& ... &
\lambda_{1} \phi_{1,n-N+1}  \\
 \phi_{N,n-N} & \phi_{N-1,n-N}& ... & \phi_{1,n-N}  \\
\end{vmatrix}
  \end{array}_{N \times N}
\end{equation}

\begin{equation}\label{Darboux_C21}
\begin{array}{ll} & W(\phi_n, \phi_{N,n}, \phi_{N-1,n}, ..., \phi_{1,n})\\
&=\begin{vmatrix}
 \lambda^N \phi_n & \lambda^{N}_N \phi_{N,n} &
\lambda^{N}_{N-1} \phi_{N-1,n}& ... &
\lambda^{N}_{1} \phi_{1,n}  \\
 \lambda^{N-1} \phi_{n-1} & \lambda^{N-1}_N \phi_{N,n-1} &
\lambda^{N-1}_{N-1} \phi_{N-1,n-1}& ... &
\lambda^{N-1}_{1} \phi_{1,n-1}  \\
 ... & ... & ... & ...  \\
 \lambda \phi_{n-N+1} & \lambda_N \phi_{N,n-N+1} & \lambda_{N-1} \phi_{N-1,n-N+1}& ... &
\lambda_{1} \phi_{1,n-N+1}  \\
\phi_{n-N}& \phi_{N,n-N} & \phi_{N-1,n-N}& ... & \phi_{1,n-N}  \\
\end{vmatrix}
  \end{array}_{(N+1) \times (N+1)}
\end{equation}

 $\phi_n[N]$ satisfies the
following linear system:
\begin{subequations}\label{volterra_lax_new}
\begin{equation}
\phi_{n+1}[N]+u_n[N] \phi_{n-1}[N]=\lambda \phi_n[N],
\end{equation}
\begin{equation}
\frac{d\phi_n[N]}{dt}=-u_n[N] \phi_{n}[N]+\lambda u_n[N]
\phi_{n-1}[N],
\end{equation}
\end{subequations}
where $u_n[N]$ are defined by
\begin{equation}\label{Darboux_C21}
u_n[N]=u_{n-N}\frac{W_{n-1}(N) W_{n+2}(N)}{W_n(N) W_{n+1}(N) };
W_n(N)=W(\phi_{N,n-1}, \phi_{N-1,n-1}, ..., \phi_{1,n-1})
\end{equation}
\end{theorem}
The proof of this theorem in the real case is given in [2].
Substituting $u_n=a_n+ib_n$ and  $
\phi(n,t)=\psi^{(1)}(n,t)+i\psi^{(2)}(n,t)$ into Theorem 2.1
 and separating the real and imaginary parts of \eqref{Darboux_C1}-\eqref{Darboux_C21}
, we obtain the following $N$-step Darboux transformation for
coupled Volterra system \eqref{Cvolterra2}:

\begin{theorem}\label{CDarboux_C}
Equation \eqref{Cvolterra_lax} are invariant with respect to Darboux
transformation

\begin{subequations}\label{CDarboux}
\begin{equation}
\psi^{(1)}_{n}[N]=\frac{W^r(\phi_{n},\phi_{N,n}, \phi_{N-1,n}, ...,
\phi_{1,n})W^r_{n}(N)+W^i(\phi_{n},\phi_{N,n}, \phi_{N-1,n}, ...,
\phi_{1,n})W^i_{n}(N)}{(W^r_{n}(N))^2+(W^i_{n}(N))^2},
\end{equation}
\begin{equation}
\psi^{(2)}_{n}[N]=\frac{W^i(\phi_{n},\phi_{N,n}, \phi_{N-1,n}, ...,
\phi_{1,n})W^r_{n}(N)-W^r(\phi_{n},\phi_{N,n}, \phi_{N-1,n}, ...,
\phi_{1,n})W^i_{n}(N)}{(W^r_{n}(N))^2+(W^i_{n}(N))^2}
\end{equation}
\end{subequations}
where $\psi^{(1)}_{i,n}$ and $\psi^{(2)}_{i,n}$ the fixed
solutions of \eqref{Cvolterra_lax} taken at the point
$\lambda=\lambda_i$ $(i=1, 2, 3, ..., N)$. $\psi^{(1)}_{n}[N]$ and
$\psi^{(2)}_{n}[N]$ satisfy the following linear systems:
\begin{subequations}
\begin{equation}
\psi^{(1)}_{n+1}[N]=\lambda \psi^{(1)}_n[N]+b_n[N]
\psi^{(2)}_{n-1}[N]-a_n[N] \psi^{(1)}_{n-1}[N],
\end{equation}
\begin{equation}
\psi^{(2)}_{n+1}[N]=\lambda \psi^{(2)}_n[N]-b_n[N]
\psi^{(1)}_{n-1}[N]-a_n[N] \psi^{(2)}_{n-1}[N],
\end{equation}
\begin{equation}
\frac{d \psi^{(1)}_{n}[N]}{d t}=b_n[N] \psi^{(2)}_n[N]-a_n[N]
\psi^{(1)}_{n}[N]+ \lambda a_n[N] \psi^{(1)}_{n-1}[N]- \lambda
b_n[N] \psi^{(2)}_{n-1}[N],
\end{equation}
\begin{equation}
\frac{d \psi^{(2)}_{n}[N]}{d t}=-b_n[N] \psi^{(1)}_n[N]-a_n[N]
\psi^{(2)}_{n}[N]+ \lambda b_n[N] \psi^{(1)}_{n-1}[N]+ \lambda
a_n[N] \psi^{(2)}_{n-1}[N],
\end{equation}
\end{subequations}
where $a_n[N]$ and $b_n[N]$ are defined by
\begin{equation}
a_n[N]=\frac{A(N)}{\Delta(N)},\qquad
b_n[N]=\frac{B(N)}{\Delta(N)},
\end{equation}
with
\begin{subequations}
\begin{equation}
\begin{array}{ll}
A(N)=&
\begin{vmatrix}
W^r_{n-1}(N) & W^i_{n+2}(N)  \\
W^i_{n-1}(N) & W^r_{n+2}(N)  \\
\end{vmatrix}
\begin{vmatrix}
W^r_n(N) & a_{n-N}W^i_{n+1}(N)-b_{n-N}W^r_{n+1}(N)  \\
W^i_n(N) & a_{n-N}W^r_{n+1}(N)+b_{n-N}W^i_{n+1}(N)  \\
\end{vmatrix} \\
&+
\begin{vmatrix}
W^r_{n-1}(N) & -W^r_{n+2}(N)  \\
W^i_{n-1}(N) & W^i_{n+2}(N)  \\
\end{vmatrix}
\begin{vmatrix}
W^r_n(N) & -b_{n-N}W^i_{n+1}(N)-a_{n-N}W^r_{n+1}(N)  \\
W^i_n(N) & -b_{n-N}W^r_{n+1}(N)+a_{n-N}W^i_{n+1}(N)  \\
\end{vmatrix}
\end{array}
\end{equation}
\begin{equation}
\begin{array}{ll}
B(N)=&
\begin{vmatrix}
W^r_{n-1}(N) & W^i_{n+2}(N)  \\
W^i_{n-1}(N) & W^r_{n+2}(N)  \\
\end{vmatrix}
\begin{vmatrix}
W^r_n(N) & b_{n-N}W^i_{n+1}(N)+a_{n-N}W^r_{n+1}(N)  \\
W^i_n(N) & b_{n-N}W^r_{n+1}(N)-a_{n-N}W^i_{n+1}(N)  \\
\end{vmatrix}\\
&+
\begin{vmatrix}
W^r_{n-1}(N) & -W^r_{n+2}(N)  \\
W^i_{n-1}(N) & W^i_{n+2}(N)  \\
\end{vmatrix}
\begin{vmatrix}
W^r_n(N) & a_{n-N}W^i_{n+1}(N)-b_{n-N}W^r_{n+1}(N)  \\
W^i_n(N) & a_{n-N}W^r_{n+1}(N)+b_{n-N}W^i_{n+1}(N)  \\
\end{vmatrix}
\end{array}
\end{equation}
\begin{equation}
\Delta(N)=\begin{vmatrix}
W^r_n(N) & W^i_{n+1}(N)  \\
W^i_n(N) & W^r_{n+1}(N)  \\
\end{vmatrix}^2
+\begin{vmatrix}
W^r_n(N) & -W^r_{n+1}(N)  \\
W^i_n(N) & W^i_{n+1}(N)  \\
\end{vmatrix}^2
\end{equation}
\begin{equation}
W^r_n(N)=\frac{W_{n}(N)+{W_n^{*}(N)}}{2},~~~~~~~~~W^i_n(N)=\frac{W_{n}(N)-{W_{n}^{*}(N)}}{2i}.
\end{equation}
\end{subequations}
Here ${W}_{n}^{*}(N)$ is conjugation of $W_{n}(N)$.
\end{theorem}

It is obvious that the solutions $ a_n[N]$ and $b_n[N]$ are
described by the wave functions of the spectral problem
\eqref{Cvolterra_lax} with $\lambda=\lambda_i$ $(i=1, 2, 3, ...,
N)$. For example, suppose $a_n, b_n$ are seed solutions, the
first-step Darboux transformation yields new solutions $a_n[1]$
and $b_n[1]$:
\begin{subequations}
\begin{equation}
a_n[1]=\frac{A(1)}{\Delta(1)},\qquad b_n[1]=\frac{B(1)}{\Delta(1)},
\end{equation}
\begin{equation}
\begin{array}{ll}
 A(1)=
   & \begin{vmatrix}
\psi^{(1)}_{1,n-2} & \psi^{(2)}_{1,n+1}  \\
\psi^{(2)}_{1,n-2} & \psi^{(1)}_{1,n+1}  \\
\end{vmatrix}
\begin{vmatrix}
\psi^{(1)}_{1,n-1} & a_{n-1}\psi^{(2)}_{1,n}-b_{n-1}\psi^{(1)}_{1,n}  \\
\psi^{(2)}_{1,n-1} & a_{n-1}\psi^{(1)}_{1,n}+b_{n-1}\psi^{(2)}_{1,n}  \\
\end{vmatrix} \\
&+\begin{vmatrix}
\psi^{(1)}_{1,n-2} & -\psi^{(1)}_{1,n+1}  \\
\psi^{(2)}_{1,n-2} & \psi^{(2)}_{1,n+1}  \\
\end{vmatrix}
\begin{vmatrix}
\psi^{(1)}_{1,n-1} & -b_{n-1}\psi^{(2)}_{1,n}-a_{n-1}\psi^{(1)}_{1,n}  \\
\psi^{(2)}_{1,n-1} & -b_{n-1}\psi^{(1)}_{1,n}+a_{n-1}\psi^{(2)}_{1,n}  \\
\end{vmatrix}, \\
\end{array}
\end{equation}
\begin{equation}
\begin{array}{ll}
 B(1)=
   &\begin{vmatrix}
\psi^{(1)}_{1,n-2} & \psi^{(2)}_{1,n+1}  \\
\psi^{(2)}_{1,n-2} & \psi^{(1)}_{1,n+1}  \\
\end{vmatrix}
\begin{vmatrix}
\psi^{(1)}_{1,n-1} & b_{n-1}\psi^{(2)}_{1,n}+a_{n-1}\psi^{(1)}_{1,n}  \\
\psi^{(2)}_{1,n-1} & b_{n-1}\psi^{(1)}_{1,n}-a_{n-1}\psi^{(2)}_{1,n}  \\
\end{vmatrix}\\
&+\begin{vmatrix}
\psi^{(1)}_{1,n-2}  & -\psi^{(1)}_{1,n+1}  \\
\psi^{(2)}_{1,n-2}  & \psi^{(2)}_{1,n+1}  \\
\end{vmatrix}
\begin{vmatrix}
\psi^{(1)}_{1,n-1} & a_{n-1}\psi^{(2)}_{1,n}-b_{n-1}\psi^{(1)}_{1,n}  \\
\psi^{(2)}_{1,n-1} & a_{n-1}\psi^{(1)}_{1,n}+b_{n-1}\psi^{(2)}_{1,n}  \\
\end{vmatrix},
\\
\end{array}
\end{equation}
\begin{equation}
\begin{array}{ll}
\Delta(1)=&\begin{vmatrix}
\psi^{(1)}_{1,n-1} & \psi^{(2)}_{1,n}  \\
\psi^{(2)}_{1,n-1} & \psi^{(1)}_{1,n} \\
\end{vmatrix}^2
+\begin{vmatrix}
\psi^{(1)}_{1,n-1} & -\psi^{(1)}_{1,n}  \\
\psi^{(2)}_{1,n-1} & \psi^{(2)}_{1,n}  \\
\end{vmatrix}^2.
  \end{array}
\end{equation}
\end{subequations}
The new wave functions are
\begin{subequations}
\begin{equation}
\psi^{(1)}_{n}[1]=\frac{W^r(\phi_{n},\phi_{1,n})\psi^{(1)}_{1,n-1}+W^i(\phi_{n},
\phi_{1,n})\psi^{(2)}_{1,n-1}}{(\psi^{(1)}_{1,n-1})^2+(\psi^{(2)}_{1,n-1})^2}
\end{equation}
\begin{equation}
\psi^{(2)}_{n}[1]=\frac{W^i(\phi_{n},\phi_{1,n})\psi^{(1)}_{1,n-1}-W^r(\phi_{n},
\phi_{1,n})\psi^{(2)}_{1,n-1}}{(\psi^{(1)}_{1,n-1})^2+(\psi^{(2)}_{1,n-1})^2}
\end{equation}
\end{subequations}
where
\begin{subequations}
\begin{equation}
W^r(\phi_{n},\phi_{1,n})=\begin{vmatrix}
\lambda \psi^{(1)}_{n} & \lambda_1 \psi^{(1)}_{1,n}  \\
 \psi^{(1)}_{n-1} &  \psi^{(1)}_{1,n-1} \\
\end{vmatrix}-\begin{vmatrix}
\lambda \psi^{(2)}_{n} & \lambda_1 \psi^{(2)}_{1,n}  \\
 \psi^{(2)}_{n-1} &  \psi^{(2)}_{1,n-1} \\
\end{vmatrix}
\end{equation}
\begin{equation}
W^i(\phi_{n},\phi_{1,n})=\begin{vmatrix}
\lambda \psi^{(2)}_{n} & \lambda_1 \psi^{(1)}_{1,n}  \\
 \psi^{(2)}_{n-1} &  \psi^{(1)}_{1,n-1} \\
\end{vmatrix}-\begin{vmatrix}
\lambda \psi^{(1)}_{n} & \lambda_1 \psi^{(2)}_{1,n}  \\
 \psi^{(1)}_{n-1} &  \psi^{(2)}_{1,n-1} \\
\end{vmatrix}
\end{equation}
\end{subequations}
and $\psi^{(1)}_{1,n}$ and $\psi^{(2)}_{1,n}$ are the solutions of
the spectral problem \eqref{Cvolterra_lax} with
$\lambda=\lambda_1$ corresponding to the seed solutions $a_n$ and
$b_n$ . The second-step Darboux transformation yields
\begin{equation}
 a_n[2]=\frac{A(2)}{\Delta(2)}, \qquad b_n[2]=\frac{B(2)}{\Delta(2)},
\end{equation}
where $A(2), B(2),$ and $\Delta(2)$ are given by equation (2.12)
with $N=2$ in which
\begin{subequations}
\begin{equation}
W^r_n(2)=\begin{vmatrix}
\lambda_2 \psi^{(1)}_{2,n-1} & \lambda_1 \psi^{(1)}_{1,n-1}  \\
 \psi^{(1)}_{2,n-2} &  \psi^{(1)}_{1,n-2}\end{vmatrix}-\begin{vmatrix}
\lambda_2 \psi^{(2)}_{2,n-1} & \lambda_1 \psi^{(2)}_{1,n-1}  \\
 \psi^{(2)}_{2,n-2} &  \psi^{(2)}_{1,n-2} \\
\end{vmatrix}
\end{equation}
\begin{equation}
W^i_n(2)=\begin{vmatrix}
\lambda_2 \psi^{(1)}_{2,n-1} & \lambda_1 \psi^{(2)}_{1,n-1}  \\
 \psi^{(1)}_{2,n-2} &  \psi^{(2)}_{1,n-2}\end{vmatrix}+\begin{vmatrix}
\lambda_2 \psi^{(2)}_{2,n-1} & \lambda_1 \psi^{(1)}_{1,n-1}  \\
 \psi^{(2)}_{2,n-2} &  \psi^{(1)}_{1,n-2} \\
\end{vmatrix}
\end{equation}
\end{subequations}
and $\psi^{(1)}_{i,n}$ and $ \psi^{(2)}_{i,n}$ are solutions of the spectral problem
\eqref{Cvolterra_lax} with $\lambda=\lambda_i(i=1,2)$.
 \setcounter{equation}{0}
\section{Multi-soliton, multi-positon, multi-negaton, and multi-periodic solutions to the coupled Volterra system (1.5)}

~~~~~~ In this section, by using the Darboux transformation, we
will construct explicit solutions for the coupled Volterra system
\eqref{Cvolterra2}. Obviously, equation (1.5) has a seed solution
$a_n=1, b_n=0$, which is related to the spectral equation
\begin{subequations}\label{}
\begin{equation}
\psi^{(i)}_{n+1}=\lambda \psi^{(i)}_n-\psi^{(i)}_{n-1}, \qquad
i=1,2
\end{equation}
\begin{equation}
\frac{d \psi^{(i)}_{n}}{d t}=\lambda
\psi^{(i)}_{n-1}-\psi^{(i)}_{n}, \qquad i=1,2.
\end{equation}
\end{subequations}
We solve this spectral equation as\\
(a) for $2<\lambda<+\infty$ or $-\infty<\lambda<-2$ ,
\begin{equation}
\psi^{(i)}_n=c_1^{(i)}
k^{-n}e^{k^2t}+c_2^{(i)}k^{n}e^{k^{-2}t},\qquad i=1,2,
\end{equation}
where $\lambda=k+\frac{1}{k}$;\\
(b) for $-2<\lambda<2$,
\begin{equation}
\psi^{(i)}_n=e^{t\cos2k}[c_1^{(i)}\cos(kn-t\sin2k)+c_2^{(i)}\sin(kn-t\sin2k)],\qquad
i=1,2,
\end{equation}
where $\lambda=2\cos(k)$; \\
(c)for $\lambda=\pm 2$,
\begin{equation}
\psi^{(i)}_n=(\pm 1)^n [c_1^{(i)}(n-2t)+c_2^{(i)}]e^t,\qquad
i=1,2.
\end{equation}
By using the Darboux transformation, we obtain the various
explicit solutions to coupled Volterra system (1.5) including
multi-soliton, multi-positon, multi-negaton, multi-periodic,
soliton-periodic,
soliton-rational, and periodic-rational solutions.\\

{\bf Example 1} Taking eigenfunctions $\psi^{(1)}_{1,n}$ and
$\psi^{(2)}_{1,n}$ as the following,
\begin{equation}
\psi^{(i)}_{1,n}=c_{11}^{(i)}
k_1^{-n}e^{k_1^2t}+c_{21}^{(i)}k_1^{n}e^{k_1^{-2}t},\qquad i=1,2;
\end{equation}
\begin{equation}
\psi^{(i)}_{1,n}=e^{t\cos2k_1}[c_{11}^{(i)}\cos(k_1
n-t\sin2k_1)+c_{21}^{(i)}\sin(k_1 n-t\sin2k_1),\qquad i=1,2;
\end{equation}
\begin{equation}
\psi^{(i)}_{1,n}=(-1)^n
[c_{11}^{(i)}(n-2t)+c_{21}^{(i)}]e^t,\qquad i=1,2,
\end{equation}
and then using 1-step Darboux transformation, we obtain 1-soliton,
1-periodic, and 1-rational solutions respectively. These solutions
are given by
\begin{subequations}
\begin{equation}
a_n[1]= \frac{ \begin{vmatrix}
\psi^{(1)}_{1,n-2} & \psi^{(2)}_{1,n+1}  \\
\psi^{(2)}_{1,n-2} & \psi^{(1)}_{1,n+1}  \\
\end{vmatrix}
\begin{vmatrix}
\psi^{(1)}_{1,n-1} & \psi^{(2)}_{1,n}  \\
\psi^{(2)}_{1,n-1} & \psi^{(1)}_{1,n}  \\
\end{vmatrix}+\begin{vmatrix}
\psi^{(1)}_{1,n-2} & -\psi^{(1)}_{1,n+1}  \\
\psi^{(2)}_{1,n-2} & \psi^{(2)}_{1,n+1}  \\
\end{vmatrix}
\begin{vmatrix}
\psi^{(1)}_{1,n-1} & -\psi^{(1)}_{1,n}  \\
\psi^{(2)}_{1,n-1} & \psi^{(2)}_{1,n}  \\
\end{vmatrix} }{\begin{vmatrix}
\psi^{(1)}_{1,n-1} & \psi^{(2)}_{1,n}  \\
\psi^{(2)}_{1,n-1} & \psi^{(1)}_{1,n}  \\
\end{vmatrix}^2
+\begin{vmatrix}
\psi^{(1)}_{1,n-1} & -\psi^{(1)}_{1,n}  \\
\psi^{(2)}_{1,n-1} & \psi^{(2)}_{1,n}  \\
\end{vmatrix}^2 }=\frac{a(1)}{\Delta(1)},
\end{equation}
\begin{equation}
b_n[1]= \frac{\begin{vmatrix}
\psi^{(1)}_{1,n-2} & \psi^{(2)}_{1,n+1}  \\
\psi^{(2)}_{1,n-2} & \psi^{(1)}_{1,n+1}  \\
\end{vmatrix}
\begin{vmatrix}
\psi^{(1)}_{1,n-1} & \psi^{(1)}_{1,n}  \\
\psi^{(2)}_{1,n-1} & -\psi^{(2)}_{1,n}  \\
\end{vmatrix}+\begin{vmatrix}
\psi^{(1)}_{1,n-2}  & -\psi^{(1)}_{1,n+1}  \\
\psi^{(2)}_{1,n-2}  & \psi^{(2)}_{1,n+1}  \\
\end{vmatrix}
\begin{vmatrix}
\psi^{(1)}_{1,n-1} & \psi^{(2)}_{1,n} \\
\psi^{(2)}_{1,n-1} & \psi^{(1)}_{1,n}  \\
\end{vmatrix} }{\begin{vmatrix}
\psi^{(1)}_{1,n-1} & \psi^{(2)}_{1,n}  \\
\psi^{(2)}_{1,n-1} & \psi^{(1)}_{1,n} \\
\end{vmatrix}^2
+\begin{vmatrix}
\psi^{(1)}_{1,n-1} & -\psi^{(1)}_{1,n}  \\
\psi^{(2)}_{1,n-1} & \psi^{(2)}_{1,n}  \\
\end{vmatrix}^2}=\frac{b(1)}{\Delta(1)}
\end{equation}
\end{subequations}
Here we write down 1-soliton solution:
\begin{eqnarray*}
a(1)&=&((c_{11}^{(1)})^2+(c_{11}^{(2)})^2)^2 k_1^{-2(2n-1)} e^{4
k_1^2 t}+((c_{21}^{(1)})^2+(c_{21}^{(2)})^2)^2 k_1^{2(2n-1)}e^{4
   k_1^{-2}
   t} \nonumber\\
   && +[2((c_{11}^{(1)})^2-(c_{11}^{(2)})^2)((c_{21}^{(1)})^2-(c_{21}^{(2)})^2)
   +8c_{11}^{(1)}c_{21}^{(1)}c_{11}^{(2)}c_{21}^{(2)}] e^{2( k_1^2+k_1^{-2})
   t}+
(k_1+k_1^{-1}+k_1^{3}+k_1^{-3})\nonumber\\
&&\times(c_{11}^{(1)}c_{21}^{(1)}+c_{11}^{(2)}c_{21}^{(2)})[((c_{11}^{(1)})^2+(c_{11}^{(2)})^2)
k_1^{-(2n-1)} e^{(3k_1^2+k_1^{-2})t}+
((c_{21}^{(1)})^2+(c_{21}^{(2)})^2) k_1^{(2n-1)}
e^{(k_1^2+3k_1^{-2})t}]
\nonumber\\
&&+
((c_{11}^{(1)})^2+(c_{11}^{(2)})^2)((c_{21}^{(1)})^2+(c_{21}^{(2)})^2)(k_1+k_1^{-1})(k_1^{3}+k_1^{-3})e^{2(
k_1^2+k_1^{-2}) t}\\
 b(1)&=
   & (c_{11}^{(1)}c_{21}^{(2)}-c_{21}^{(1)}c_{11}^{(2)})(k_1^{3}+k_1^{-3}-k_1-k_1^{-1})[((c_{11}^{(1)})^2+(c_{11}^{(2)})^2)
 k_1^{-(2n-1)}
 e^{(3k_1^2+k_1^{-2})t}\nonumber\\
 && +
((c_{21}^{(1)})^2+(c_{21}^{(2)})^2)
 k_1^{2n-1}  e^{(k_1^2+3k_1^{-2})t}]\\
\Delta(1)&=
   &[2((c_{11}^{(1)})^2-(c_{11}^{(2)})^2)((c_{21}^{(1)})^2-(c_{21}^{(2)})^2)
   +8c_{11}^{(1)}c_{21}^{(1)}c_{11}^{(2)}c_{21}^{(2)}] e^{2( k_1^2+k_1^{-2})
   t}+2(c_{11}^{(1)}c_{21}^{(1)}+c_{11}^{(2)}c_{21}^{(2)})\nonumber\\
   &&\times (k_1+k_1^{-1})[((c_{11}^{(1)})^2+(c_{11}^{(2)})^2) k_1^{-(2n-1)}
e^{(3k_1^2+k_1^{-2})t}+((c_{21}^{(1)})^2+(c_{21}^{(2)})^2)k_1^{2n-1}
e^{(k_1^2+3k_1^{-2})t}]
\end{eqnarray*}

Their plots are given in the Fig.1.(1-soliton with $c_{11}^{(1)} =
c_{21}^{(1)} = c_{11}^{(2)}= 1,c_{21}^{(2)} =-1; k_1=2$; $t=5$;
1-periodic with $c_{11}^{(1)} = c_{21}^{(1)} = c_{11}^{(2)}=
1,c_{21}^{(2)} =2; k_1=\frac{\pi}{20}$; $t=2$ and
$t=2+\frac{\pi}{\sin\frac{\pi}{10}}$; 1-rational with
$c_{11}^{(1)}
=c_{21}^{(1)} = c_{11}^{(2)}= 1, c_{21}^{(2)} =-10; \lambda_1=2$;$t=1$).\\
\begin{center}
\begin{tabular}{ccc}
\includegraphics[width=5cm]{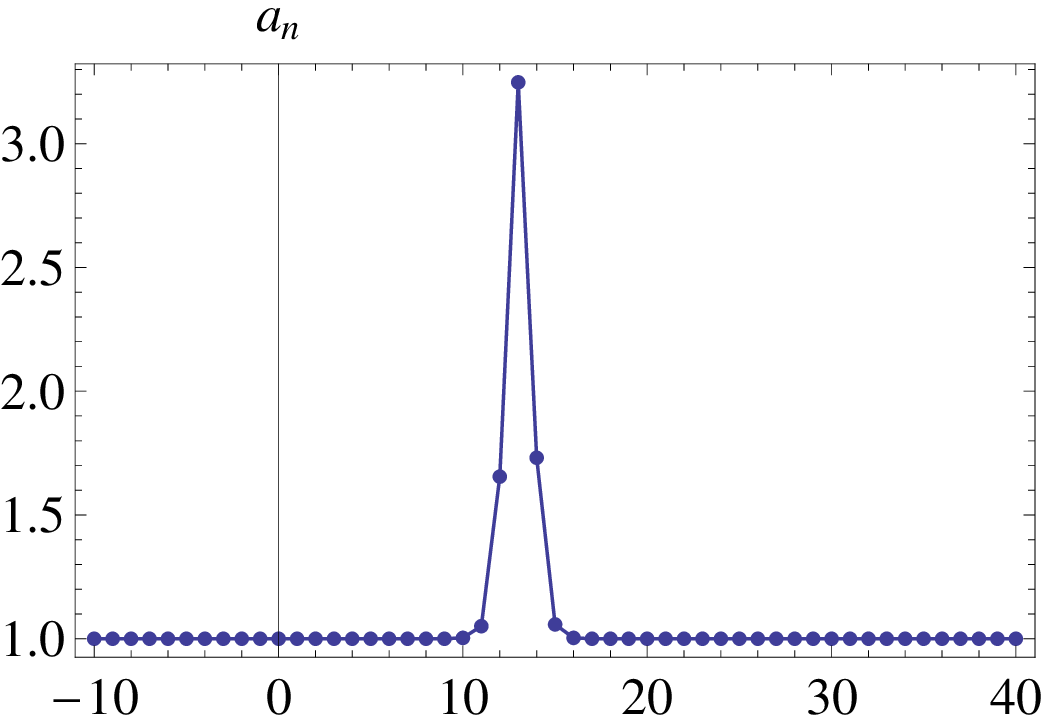} ~~
\includegraphics[width=5cm]{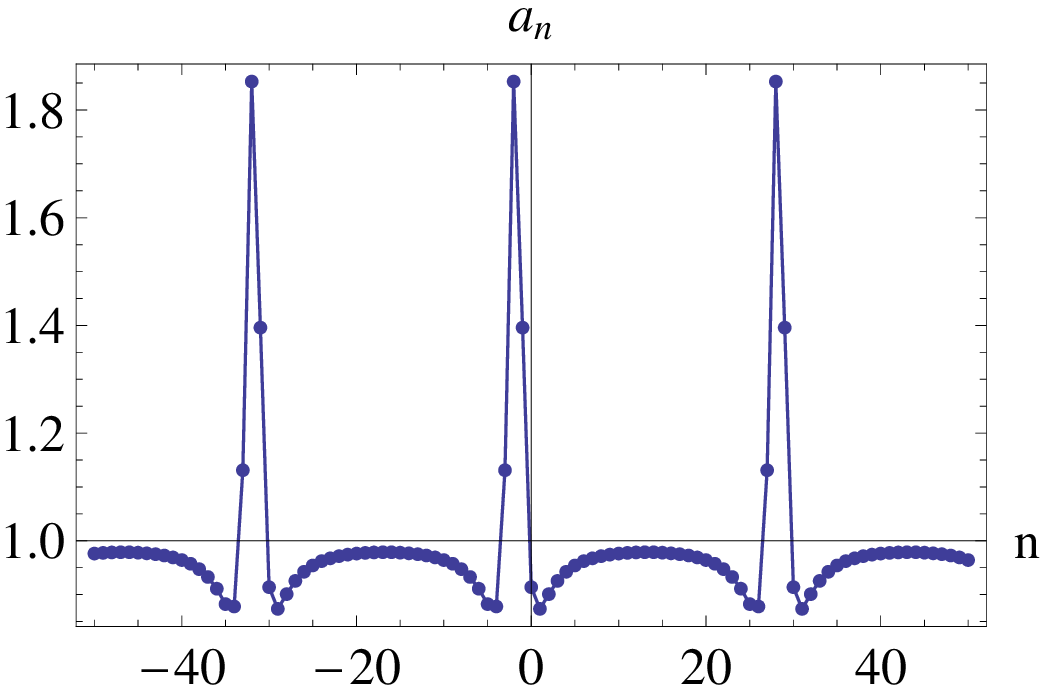}~~
\includegraphics[width=5cm]{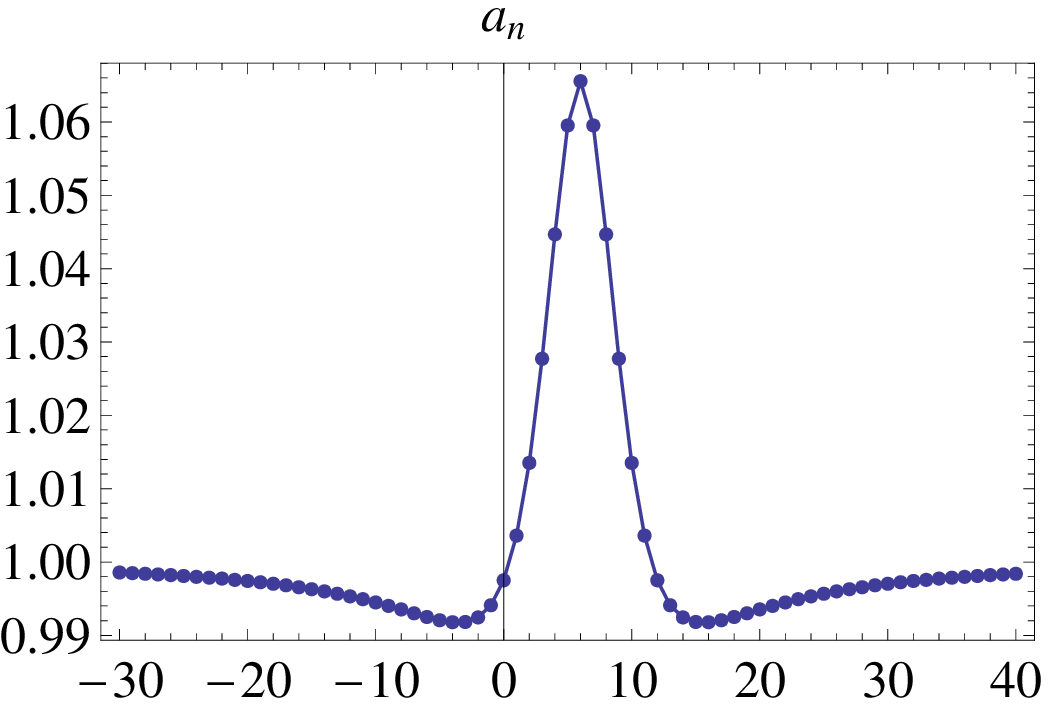} \\
\includegraphics[width=5cm]{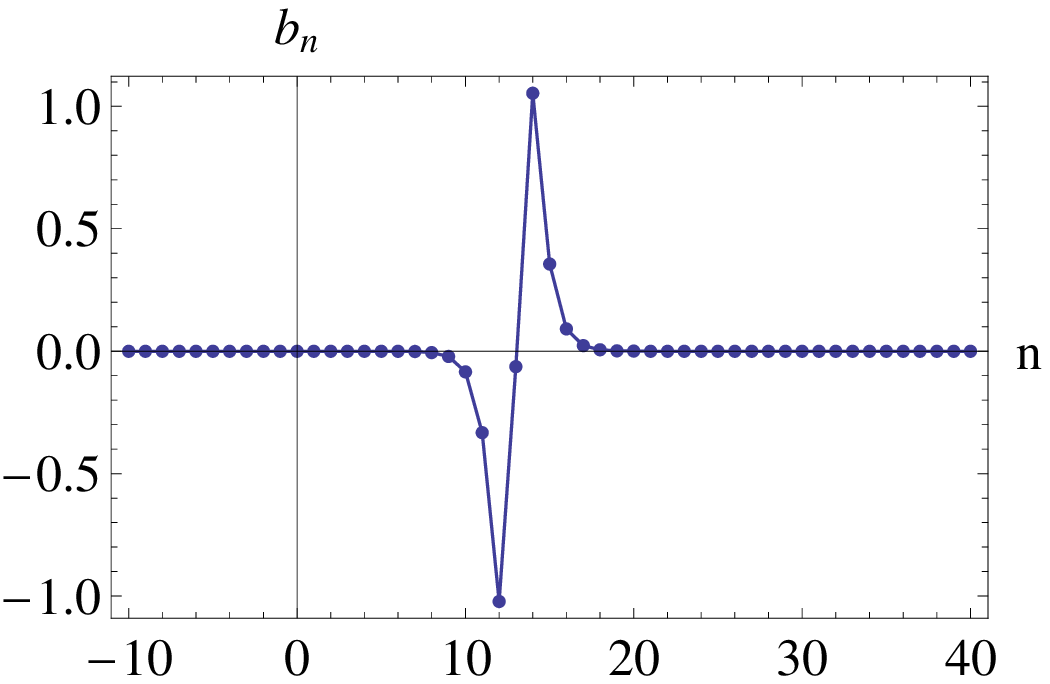} ~~
\includegraphics[width=5cm]{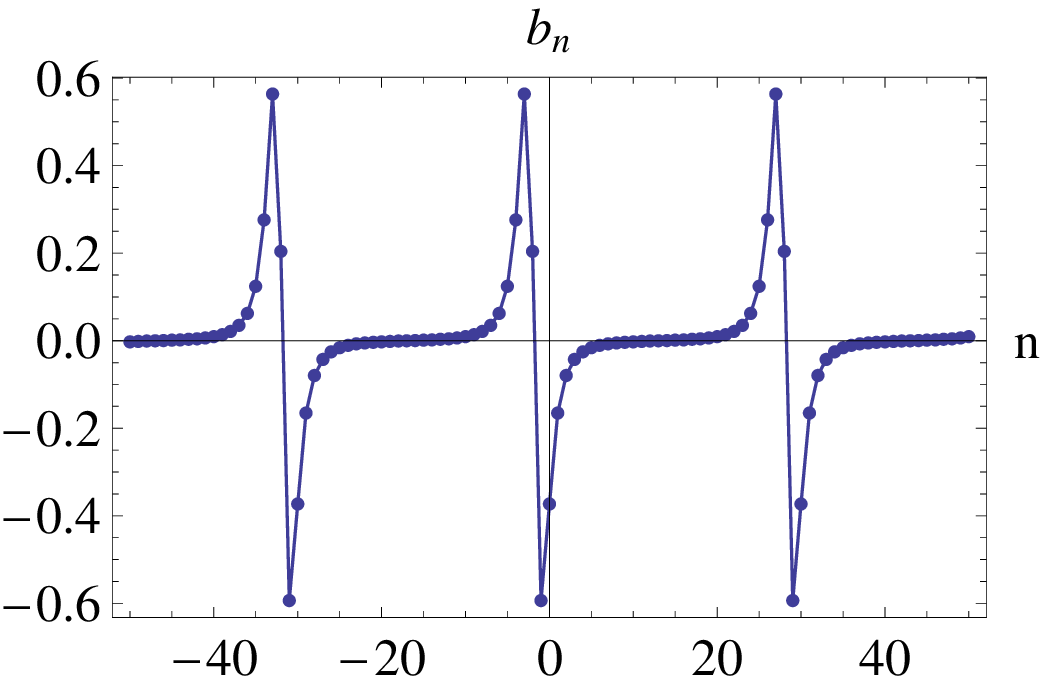}  ~~
\includegraphics[width=5cm]{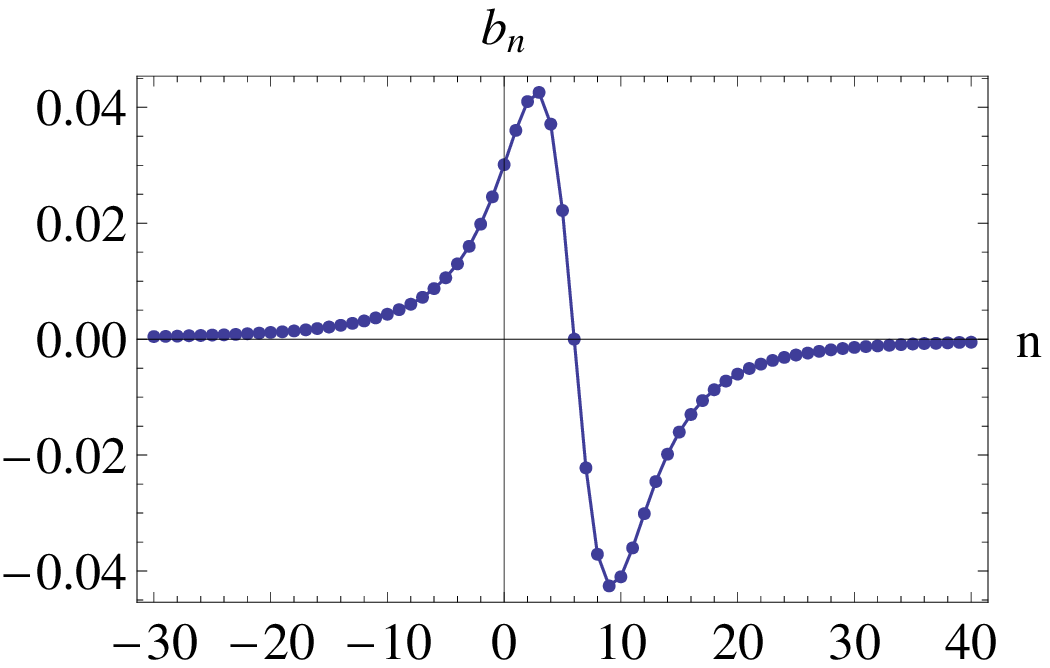}
 \\
 Fig. 1.~~ 1-soliton, ~~~ 1-periodic, ~~~ 1-rational
\end{tabular}
\end{center}

{\bf Example 2} Taking eigenfunctions $\psi^{(1)}_{j,n}$ and
$\psi^{(2)}_{j,n} (j=1,2)$ as the following,
\begin{equation}
\psi^{(i)}_{j,n}=c_{1j}^{(i)}
k_j^{-n}e^{k_j^2t}+c_{2j}^{(i)}k_j^{n}e^{k_j^{-2}t}, (i,j=1,2),
\end{equation}
\begin{equation}
\psi^{(i)}_{j,n}=e^{t\cos2k_j}[c_{1j}^{(i)}\cos(k_j
n-t\sin2k_j)+c_{2j}^{(i)}\cos(k_j n-t\sin2k_j)],(i,j=1,2),
\end{equation}
we obtain 2-soliton and 2-periodic solutions, respectively. Taking
$\psi_{1,n}^{(i)}$ and $\psi_{2,n}^{(i)}$ (i=1,2) as the forms
(3.5) and (3.6),
 or (3.5) and (3.7), or (3.6) and (3.7),
 we obtain soliton-periodic, soliton-rational and periodic-rational solutions,
 respectively. They are given by
$$ a_n[2]= \frac{ \begin{vmatrix}
W^r_{n-1}(2) & W^i_{n+2}(2)  \\
W^i_{n-1}(2) & W^r_{n+2}(2)  \\
\end{vmatrix}
\begin{vmatrix}
W^r_n(2) & W^i_{n+1}(2)  \\
W^i_n(2) & W^r_{n+1}(2)  \\
\end{vmatrix} +\begin{vmatrix}
W^r_{n-1}(2) & -W^r_{n+2}(2)  \\
W^i_{n-1}(2) & W^i_{n+2}(2)  \\
\end{vmatrix}
\begin{vmatrix}
W^r_n(2) & -W^r_{n+1}(2)  \\
W^i_n(2) & W^i_{n+1}(2)  \\
\end{vmatrix} }{\begin{vmatrix}
W^r_n(2) & W^i_{n+1}(2)  \\
W^i_n(2) & W^r_{n+1}(2)  \\
\end{vmatrix}^2
+\begin{vmatrix}
W^r_n(2) & -W^r_{n+1}(2)  \\
W^i_n(2) & W^i_{n+1}(2)  \\
\end{vmatrix}^2}$$

$$ b_n[2]= \frac{ \begin{vmatrix}
W^r_{n-1}(2) & W^i_{n+2}(2)  \\
W^i_{n-1}(2) & W^r_{n+2}(2)  \\
\end{vmatrix}
\begin{vmatrix}
W^r_n(2) & W^r_{n+1}(2)  \\
W^i_n(2) & -W^i_{n+1}(2)  \\
\end{vmatrix}+\begin{vmatrix}
W^r_{n-1}(2) & -W^r_{n+2}(2)  \\
W^i_{n-1}(2) & W^i_{n+2}(2)  \\
\end{vmatrix}
\begin{vmatrix}
W^r_n(2) & W^i_{n+1}(2)  \\
W^i_n(2) & W^r_{n+1}(2)  \\
\end{vmatrix} }{\begin{vmatrix}
W^r_n(2) & W^i_{n+1}(2)  \\
W^i_n(2) & W^r_{n+1}(2)  \\
\end{vmatrix}^2
+\begin{vmatrix}
W^r_n(2) & -W^r_{n+1}(2)  \\
W^i_n(2) & W^i_{n+1}(2)  \\
\end{vmatrix}^2}$$
where
\begin{subequations}
\begin{equation}
W^r_n(2)=\begin{vmatrix}
\lambda_2 \psi^{(1)}_{2,n-1} & \lambda_1 \psi^{(1)}_{1,n-1}  \\
 \psi^{(1)}_{2,n-2} &  \psi^{(1)}_{1,n-2}\end{vmatrix}-\begin{vmatrix}
\lambda_2 \psi^{(2)}_{2,n-1} & \lambda_1 \psi^{(2)}_{1,n-1}  \\
 \psi^{(2)}_{2,n-2} &  \psi^{(2)}_{1,n-2} \\
\end{vmatrix}
\end{equation}
\begin{equation}
W^i_n(2)=\begin{vmatrix}
\lambda_2 \psi^{(1)}_{2,n-1} & \lambda_1 \psi^{(2)}_{1,n-1}  \\
 \psi^{(1)}_{2,n-2} &  \psi^{(2)}_{1,n-2}\end{vmatrix}+\begin{vmatrix}
\lambda_2 \psi^{(2)}_{2,n-1} & \lambda_1 \psi^{(1)}_{1,n-1}  \\
 \psi^{(2)}_{2,n-2} &  \psi^{(1)}_{1,n-2} \\
\end{vmatrix}
\end{equation}
\end{subequations}
For the 2-soliton case, we have
\begin{subequations}
\begin{equation}
\begin{array}{ll}
W^r_n(2))=
   &(\lambda_2 k_1-\lambda_1 k_2)(c_{11}^{(1)}c_{12}^{(1)}-c_{11}^{(2)}c_{12}^{(2)})
(k_1 k_2)^{-(n-1)} e^{(k_1^2+k_2^{2})t}\\
&+(\lambda_2 k_2-\lambda_1
k_1)(c_{21}^{(1)}c_{22}^{(1)}-c_{21}^{(2)}c_{22}^{(2)}) (k_1
k_2)^{n-2} e^{(k_1^{-2}+k_2^{-2})t}\\
 &+(\lambda_2 k_1^{-1}-\lambda_1 k_2)(c_{12}^{(1)}c_{21}^{(1)}-c_{12}^{(2)}c_{21}^{(2)})
(\frac{k_1}{ k_2})^{n-1} e^{(k_1^{-2}+k_2^{2})t}
\\
&+(\lambda_2 k_1^{}-\lambda_1
k_2^{-1})(c_{11}^{(1)}c_{22}^{(1)}-c_{11}^{(2)}c_{22}^{(2)})
(\frac{k_1}{ k_2})^{-(n-1)} e^{(k_1^{2}+k_2^{-2})t}
  \end{array}
\end{equation}
\begin{equation}
\begin{array}{ll}
W^i_n(2)=
   &(\lambda_2 k_1-\lambda_1 k_2)(c_{11}^{(2)}c_{12}^{(1)}+c_{11}^{(1)}c_{12}^{(2)})
(k_1 k_2)^{-(n-1)} e^{(k_1^2+k_2^{2})t}\\
&+(\lambda_2 k_2-\lambda_1
k_1)(c_{22}^{(1)}c_{21}^{(2)}+c_{22}^{(2)}c_{21}^{(1)}) (k_1
k_2)^{n-2} e^{(k_1^{-2}+k_2^{-2})t} \\
&+(\lambda_2 k_1^{-1}-\lambda_1
k_2)(c_{12}^{(1)}c_{21}^{(2)}+c_{12}^{(2)}c_{21}^{(1)})
(\frac{k_1}{ k_2})^{n-1} e^{(k_1^{-2}+k_2^{2})t}
\\
&+(\lambda_2 k_1^{}-\lambda_1
k_2^{-1})(c_{11}^{(2)}c_{22}^{(1)}+c_{11}^{(1)}c_{22}^{(2)})
(\frac{k_1}{ k_2})^{-(n-1)} e^{(k_1^{2}+k_2^{-2})t}
  \end{array}
\end{equation}
\end{subequations}

The Fig.2 describes the evolutions of a 2-soliton with
$c_{1i}^{(1)} = c_{2i}^{(1)} = 1, c_{11}^{(2)}=c_{22}^{(2)}=
1,c_{12}^{(2)}=c_{21}^{(2)} =-1,$ $k_1=2, k_2=3$. Next let us give
an analysis of the periodic property for the 1-periodic and the
2-periodic solutions. Note that
\begin{eqnarray*}
&&a_{n}[1]=R_1(\sin\theta_1,\cos\theta_1)\qquad
b_{n}[1]=R_2(\sin\theta_1,\cos\theta_1),
\end{eqnarray*}
where $R_1,R_2$ are two rational functions of variables, and
$\theta_1=2k_1n-2t\sin2k_1$. We thus have
\begin{subequations}
\begin{eqnarray}
&&a_{n}[1]=a_{n+\frac{\pi}{k_1}}[1],\qquad
b_{n}[1]=b_{n+\frac{\pi}{k_1}}[1],\\
&&a_{n}[1](t)=a_{n}[1](t+\frac{\pi}{\sin2k_1}), \qquad
b_{n}[1](t)=b_{n}[1](t+\frac{\pi}{\sin2k_1})
\end{eqnarray}
\end{subequations}
where $\frac{\pi}{k_1}$ is an integer. This means that the
solutions $a_n[1]$ and $b_n[1]$ are periodic in both space and
time. As for 2-periodic solution case, the periodic property is
dependent on the choice of $k_1$ and $k_2$. A tedious computation
yields
\begin{equation*}
a_n[2]=R_3(\sin\theta_1,\cos\theta_1,\sin\theta_2,\cos\theta_2,\sin(\theta_1+\theta_2),\cos(\theta_1+\theta_2),\sin(\theta_1-\theta_2),\cos(\theta_1-\theta_2))\\
\end{equation*}
\begin{equation*}
b_n[2]=R_4(\sin\theta_1,\cos\theta_1,\sin\theta_2,\cos\theta_2,\sin(\theta_1+\theta_2),\cos(\theta_1+\theta_2),\sin(\theta_1-\theta_2),\cos(\theta_1-\theta_2))
\end{equation*}
where $\theta_2=2k_2n-2t\sin2k_2$, $R_3$ and $R_4$ are two
rational functions of variables. We thus obtain
\begin{subequations}
\begin{eqnarray}
&&a_{n}[2]=a_{n+\frac{m_1\pi}{k_1}}[2],~~~~
~~~~~~~b_{n}[2]=b_{n+\frac{m_1\pi}{k_1}}[2],\\
&&a_{n}[2](t)=a_{n}[2](t+\frac{m_3\pi}{\sin 2k_1}),~~~~~~~~~
b_{n}[2](t)=b_{n}[2](t+\frac{m_3\pi}{\sin 2k_1})
\end{eqnarray}
\end{subequations}
where $m_i (i=1,2,3,4)$, $\frac{m_1\pi}{k_1}$ and
$\frac{m_2\pi}{k_2}$ are positive integers, and $m_i, k_1$ and
$k_2$ satisfy the following conditions:
\begin{eqnarray}
\frac{k_1}{k_2}=\frac{m_1}{m_2},\qquad \frac{\sin k_1}{\sin
k_2}=\frac{m_3}{m_4}.
\end{eqnarray}
The 2-periodic solutions $a_n[2]$ and $b_n[2]$ therefore are
periodic in both space and time under the proper conditions. In
the Fig.3, we make the plots for two cases: (1) $a_n[2]$, $b_n[2]$
are periodic in space; (2) $a_n[2]$, $b_n[2]$ are periodic in both
space and time (two different 2-periodic solutions with
$c_{1i}^{(1)} =c_{2i}^{(1)} = c_{1i}^{(2)}= 1,c_{21}^{(2)} =2,
c_{22}^{(2)}=-2$; for the first case, $k_1=\frac{\pi}{20},
k_2=\frac{\pi}{10}$ and $t=2$; for the second case,
$k_1=\frac{5\pi}{12}, k_2=\frac{\pi}{12}$ and $t=2$ or
$t=2+2\pi$). The plots of soliton-periodic, soliton-rational, and
periodic-rational are given in the Fig. 4 (soliton-periodic with
$c_{1i}^{(1)} =c_{2i}^{(1)} =1, c_{11}^{(2)}= 1, c_{12}^{(2)}=-1,
c_{21}^{(2)} =-1,$$ c_{22}^{(2)}=3;k_1=2, k_2=3$; soliton-rational
with $c_{1i}^{(1)} =c_{2i}^{(1)} =1, c_{21}^{(2)}= 1,
c_{12}^{(2)}=-1,$$ c_{21}^{(2)} =-1, c_{22}^{(2)}=3;k_1=2,
\lambda_2=2$; periodic-rational with $c_{1i}^{(1)} =c_{2i}^{(1)}
=1, c_{11}^{(2)}= 1, c_{12}^{(2)}=-1, c_{21}^{(2)} =-1,$$
c_{22}^{(2)}=3;k_1=2, \lambda_2=2$
)\\
\begin{center}
\begin{tabular}{ccc}
\includegraphics[width=5cm]{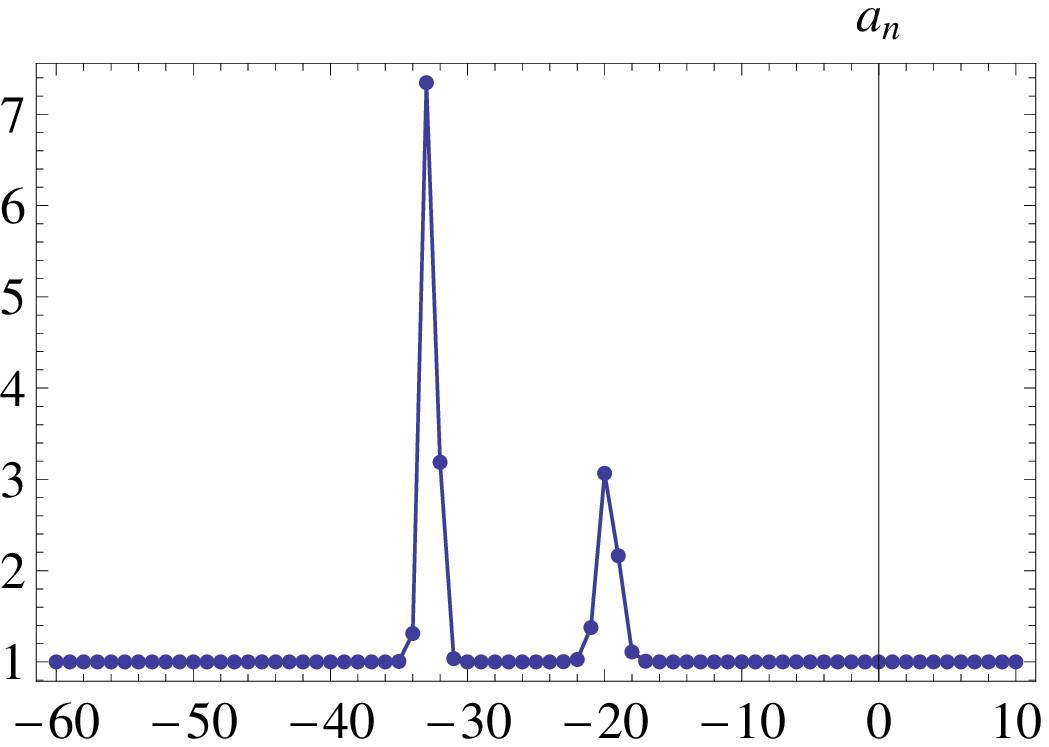} &
\includegraphics[width=5cm]{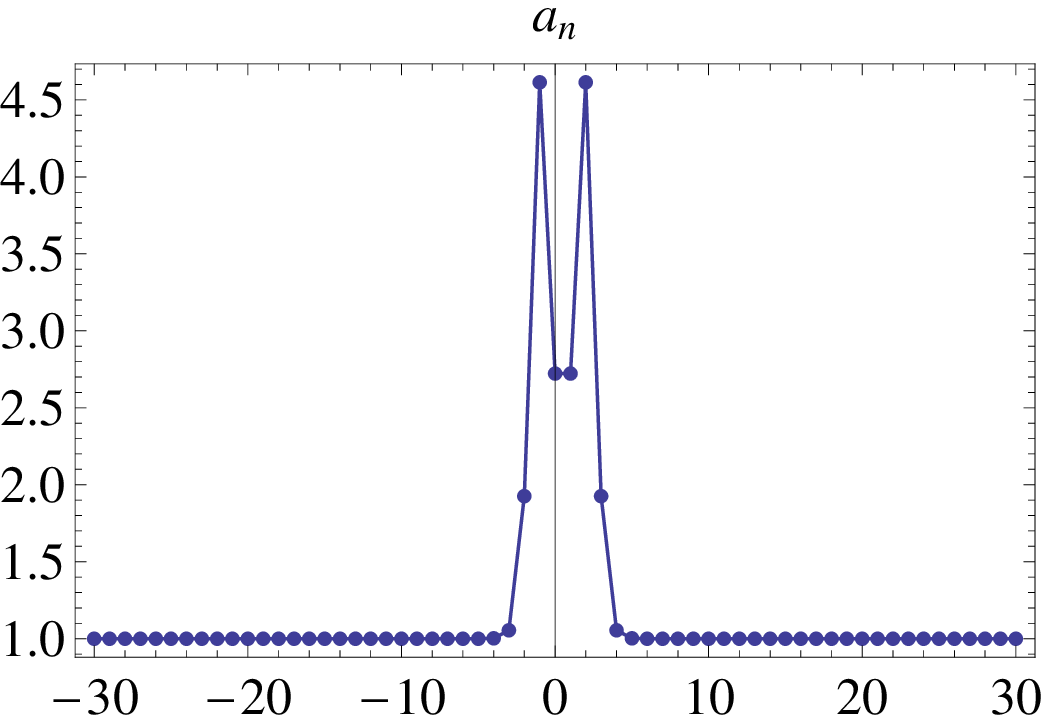} &
\includegraphics[width=5cm]{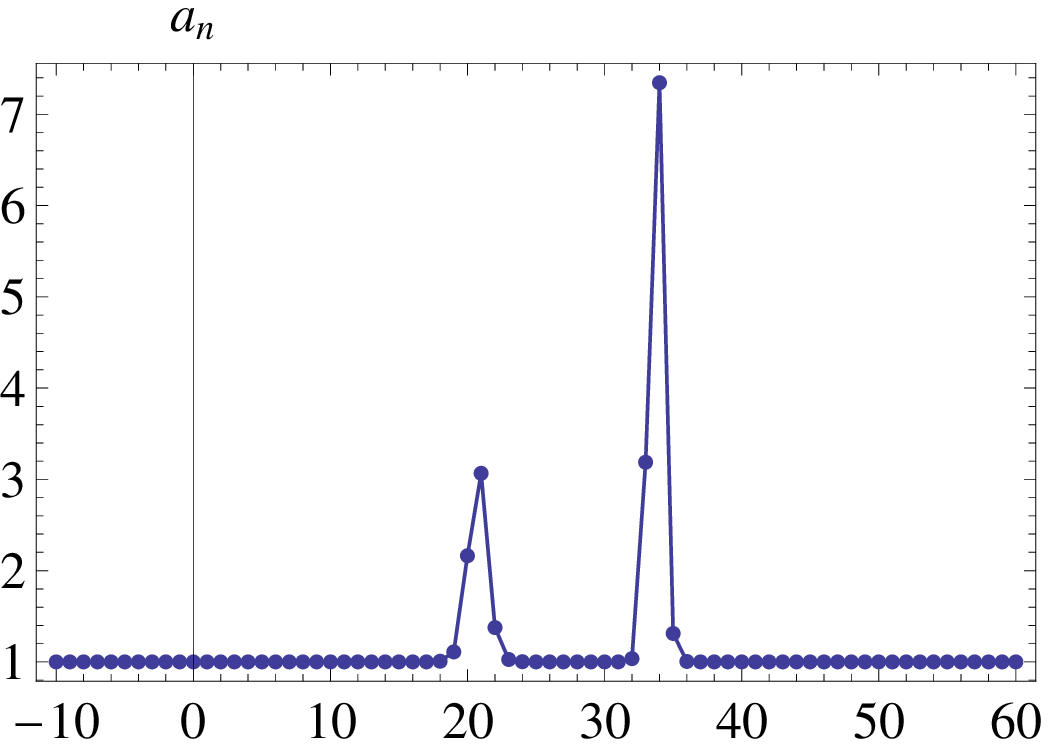} \\
  (a)  &
(b) & (c)  \\
\includegraphics[width=5cm]{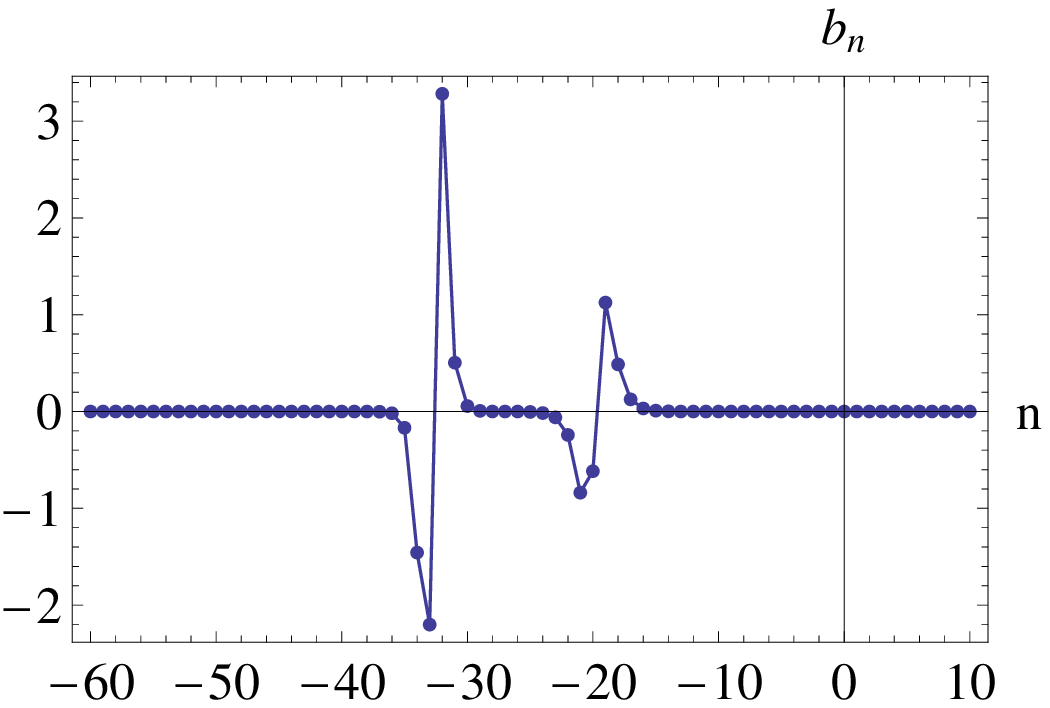} &
\includegraphics[width=5cm]{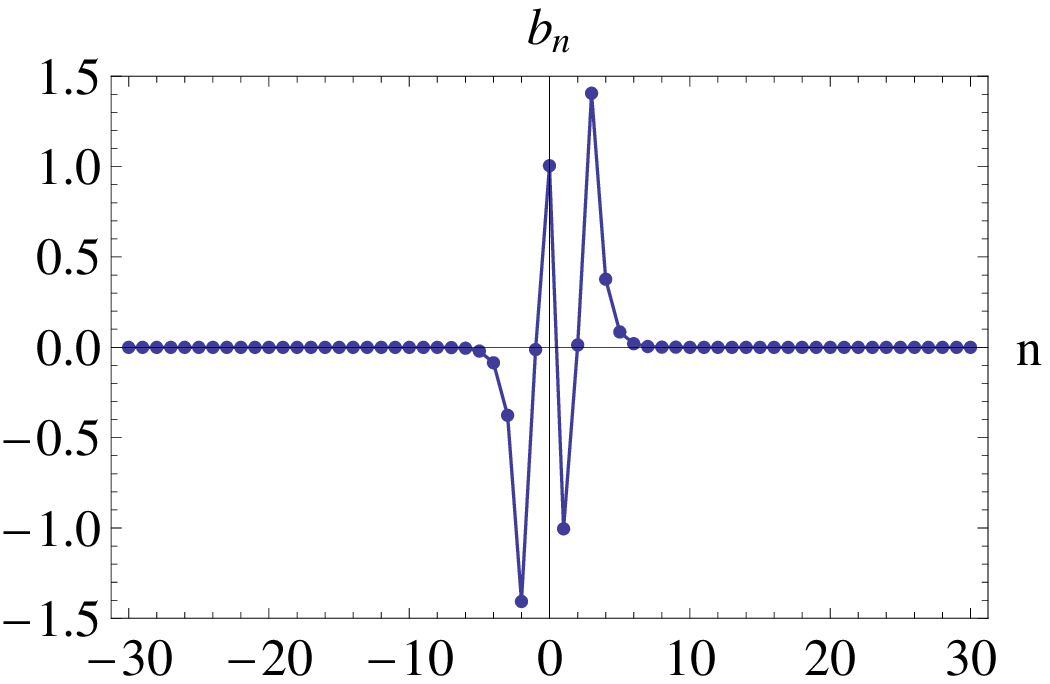} &
\includegraphics[width=5cm]{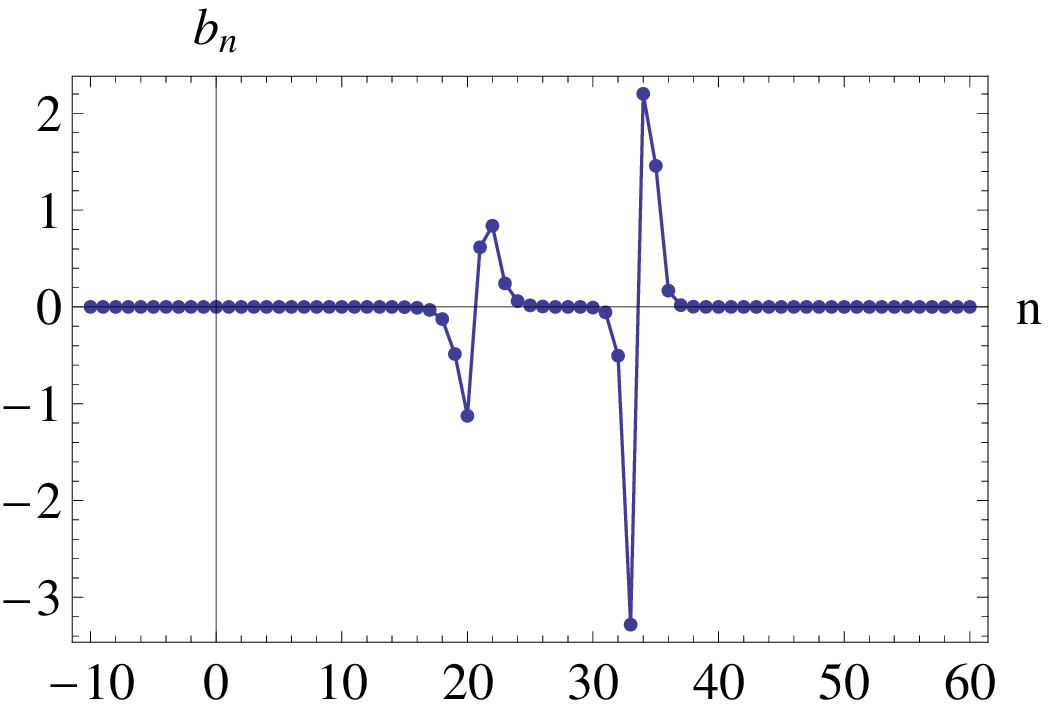} \\
(d)  & (e) & (f)   \\
\end{tabular}
 Fig. 2.~~ The evolutions of 2-soliton.
 (a),(d)
$t=-8$ ,~~~(b),(e) $t=0$ ,~~~(c),(f) $t=8$.
\end{center}

\begin{center}
\begin{tabular}{cccc}
\includegraphics[width=3.5cm]{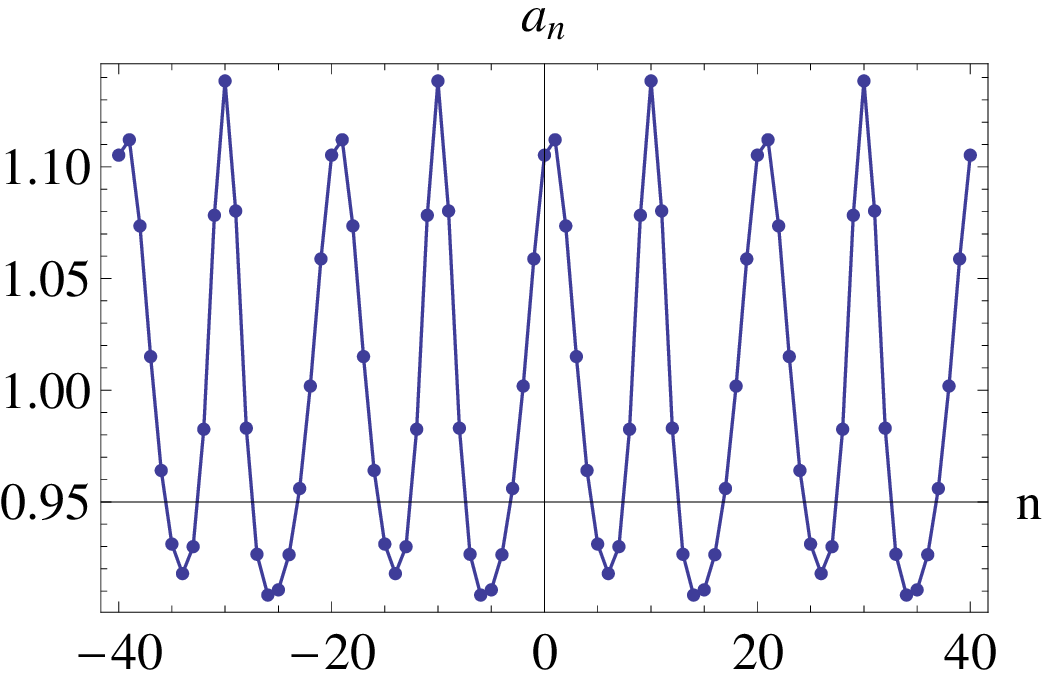} &
\includegraphics[width=3.5cm]{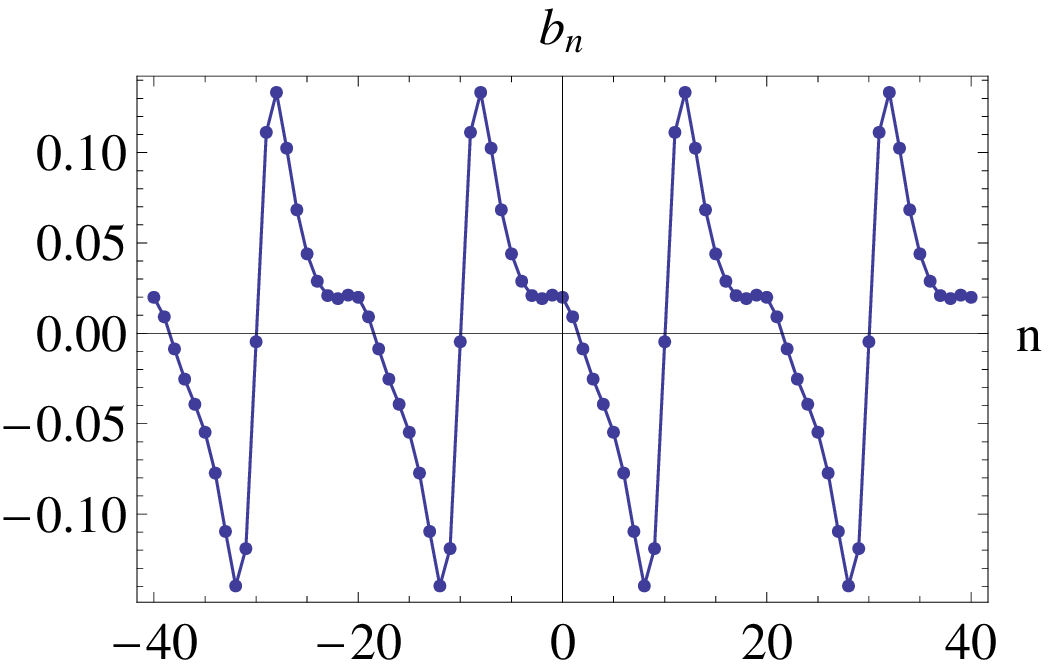} &
\includegraphics[width=3.5cm]{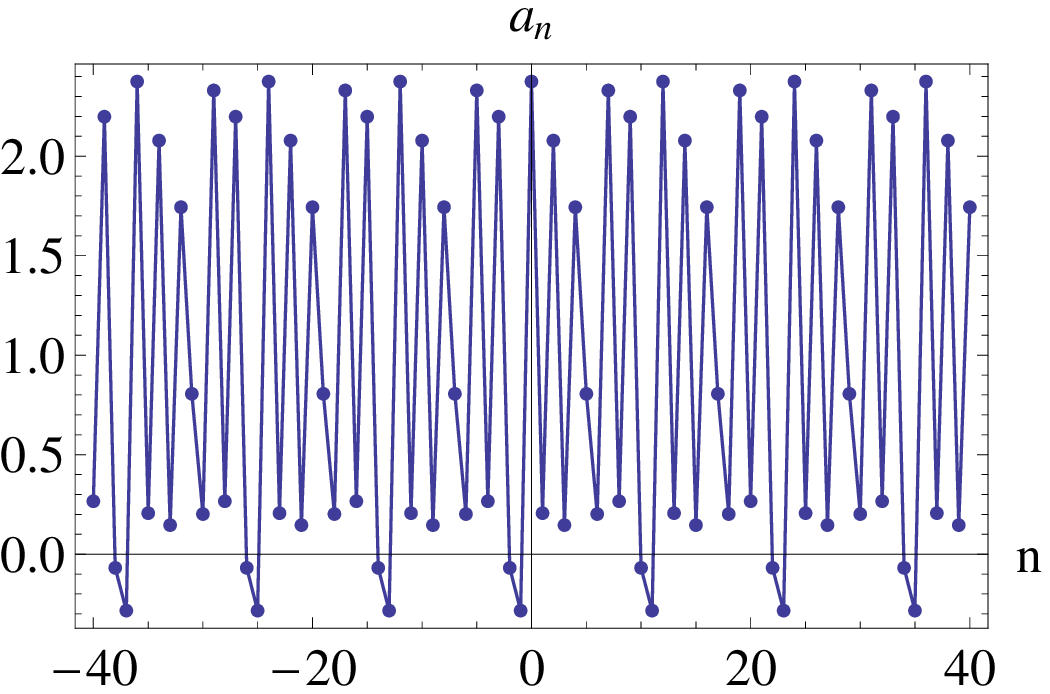}  &
\includegraphics[width=3.5cm]{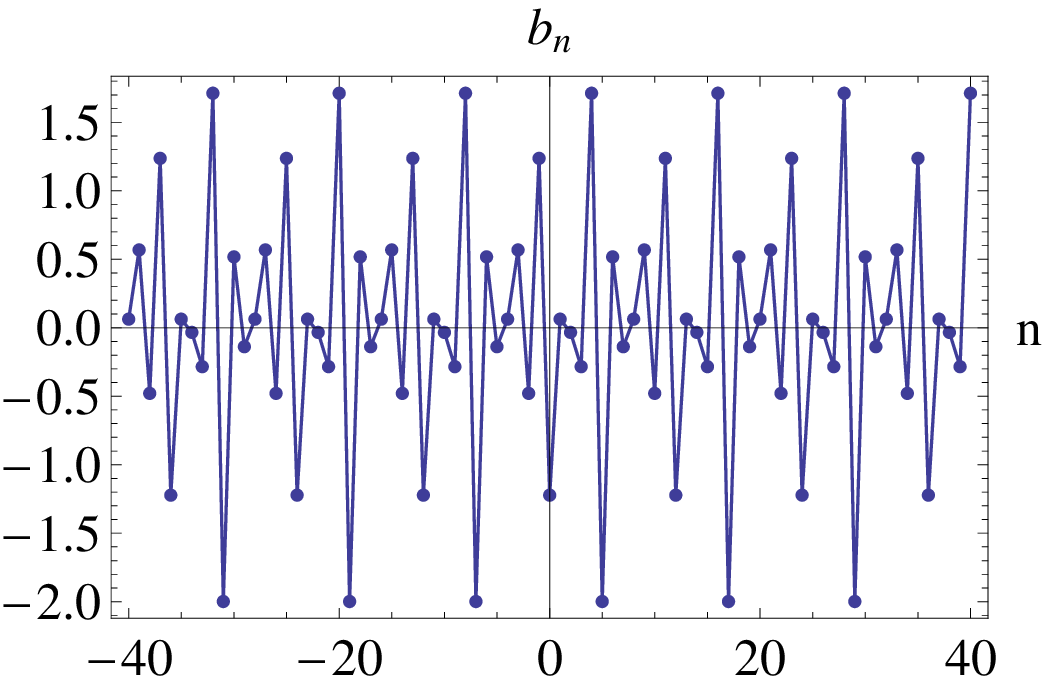} \\
  (a) & (b) &  (c) & (d) \\
\end{tabular}
 \\ Fig. 3.~~ 2-periodic solutions \\
 (a)(b) $k_1=\frac{\pi}{20}, k_2=\frac{\pi}{10} ~~ and ~~ t=2$
,~~~(c)(d) $k_1=\frac{5\pi}{12}, k_2=\frac{\pi}{12} ~~ and~~
t=2~~or ~~t=2+2\pi$
\end{center}

\begin{center}
\begin{tabular}{cccc}
\includegraphics[width=3.5cm]{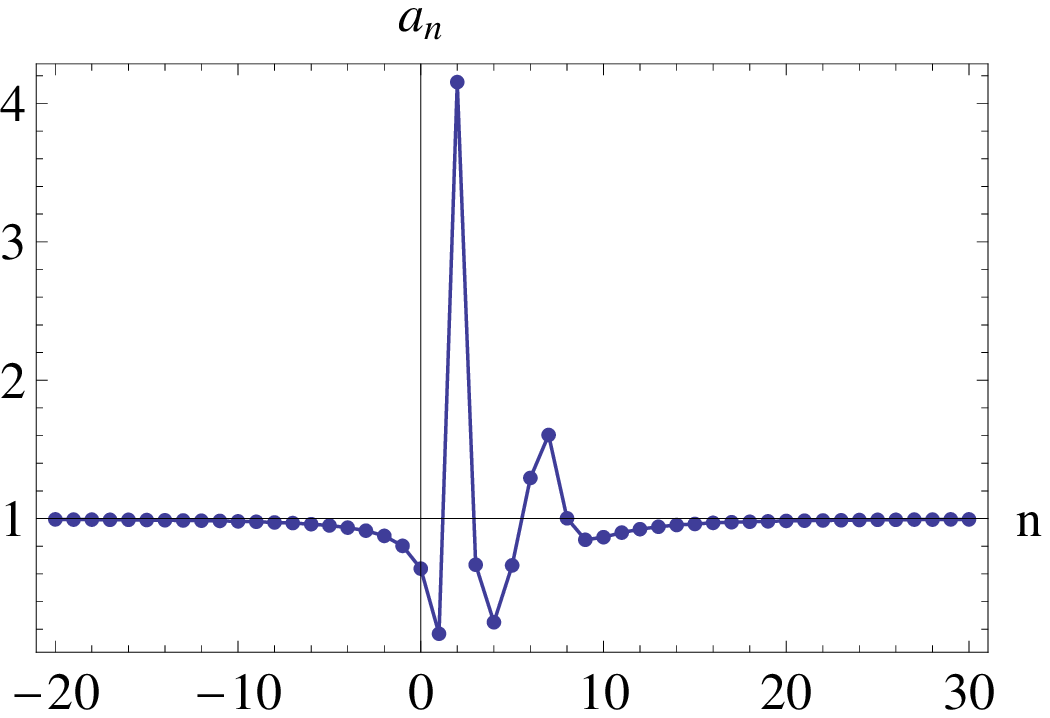} ~~
\includegraphics[width=3.5cm]{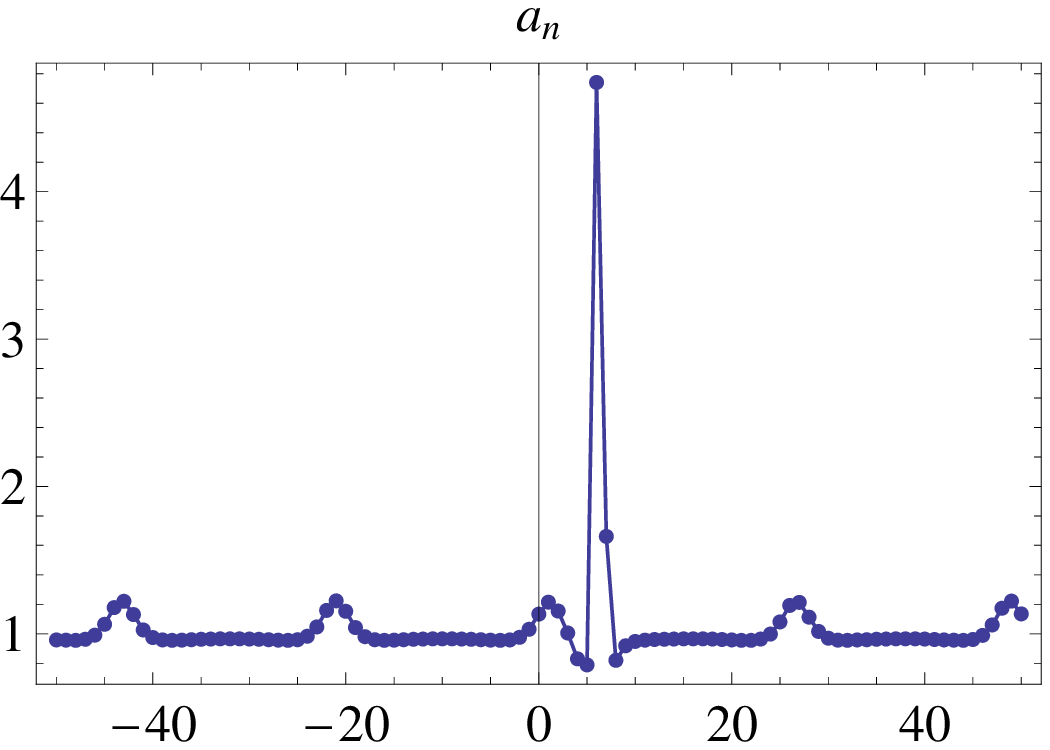} ~~
\includegraphics[width=3.5cm]{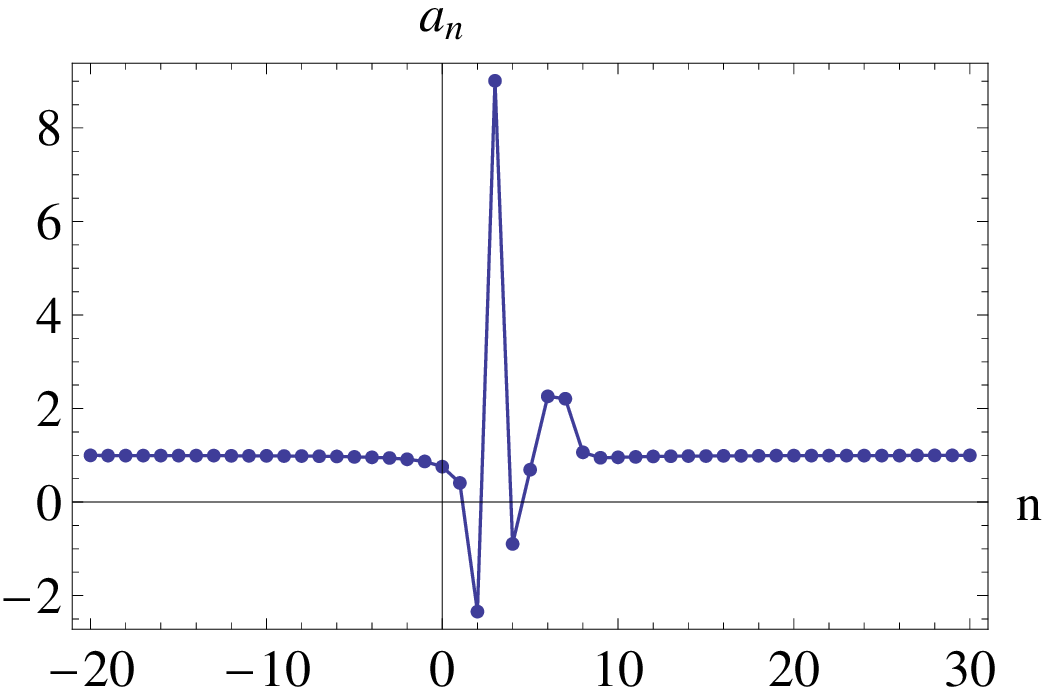} ~~
\includegraphics[width=3.5cm]{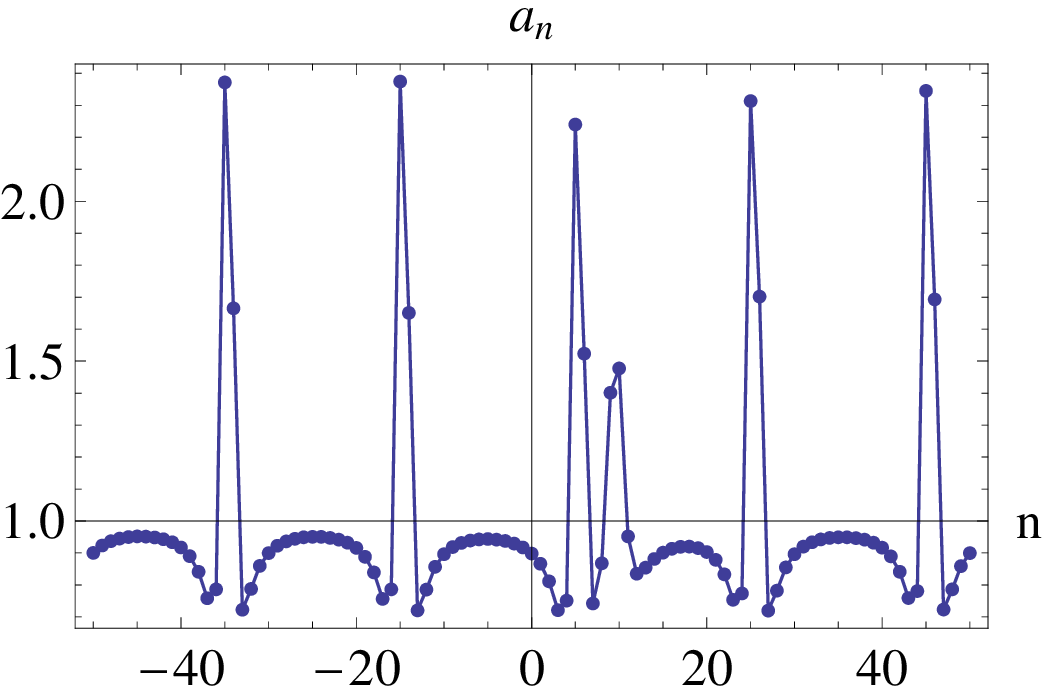}
 \\
\end{tabular}
\end{center}

\begin{center}
\begin{tabular}{ccc}
\includegraphics[width=3.5cm]{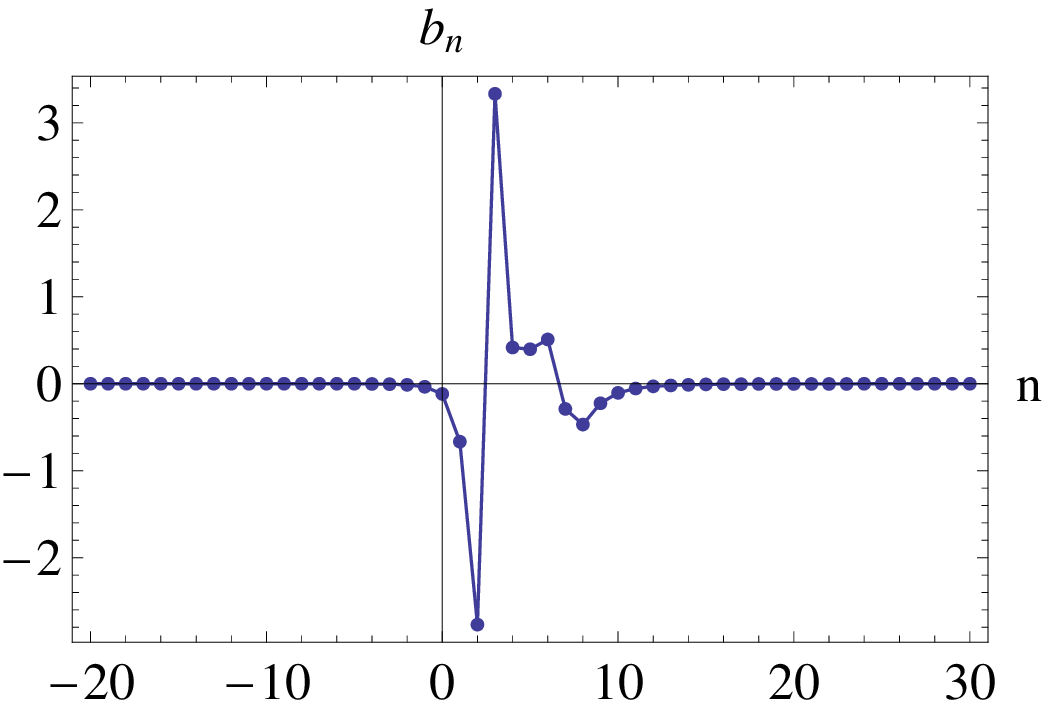}~~
\includegraphics[width=3.5cm]{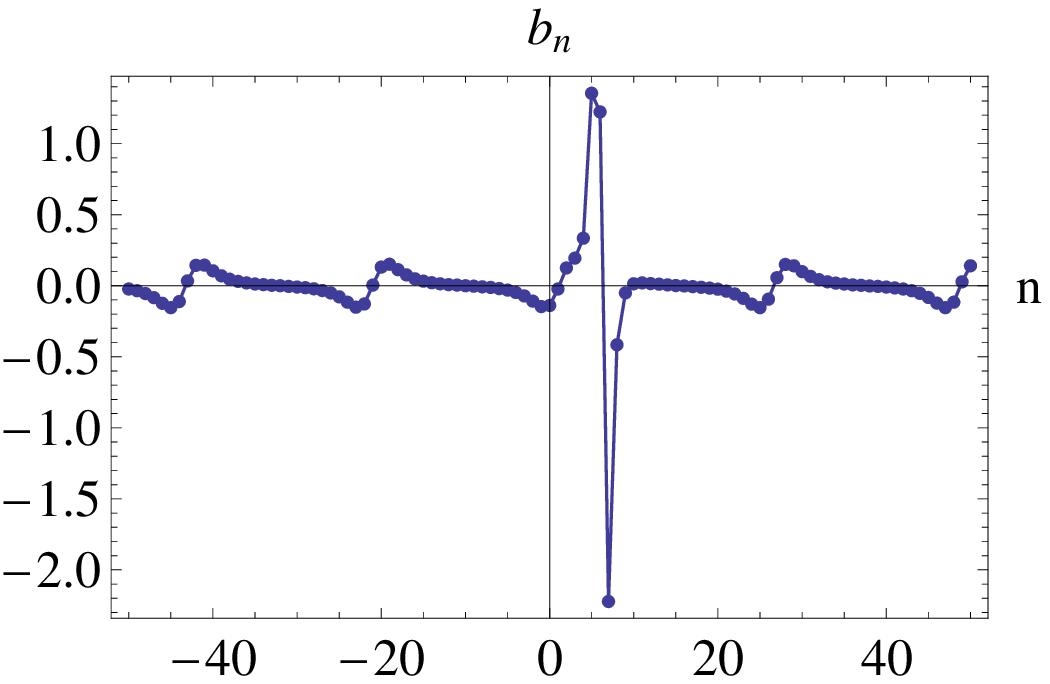} ~~
\includegraphics[width=3.5cm]{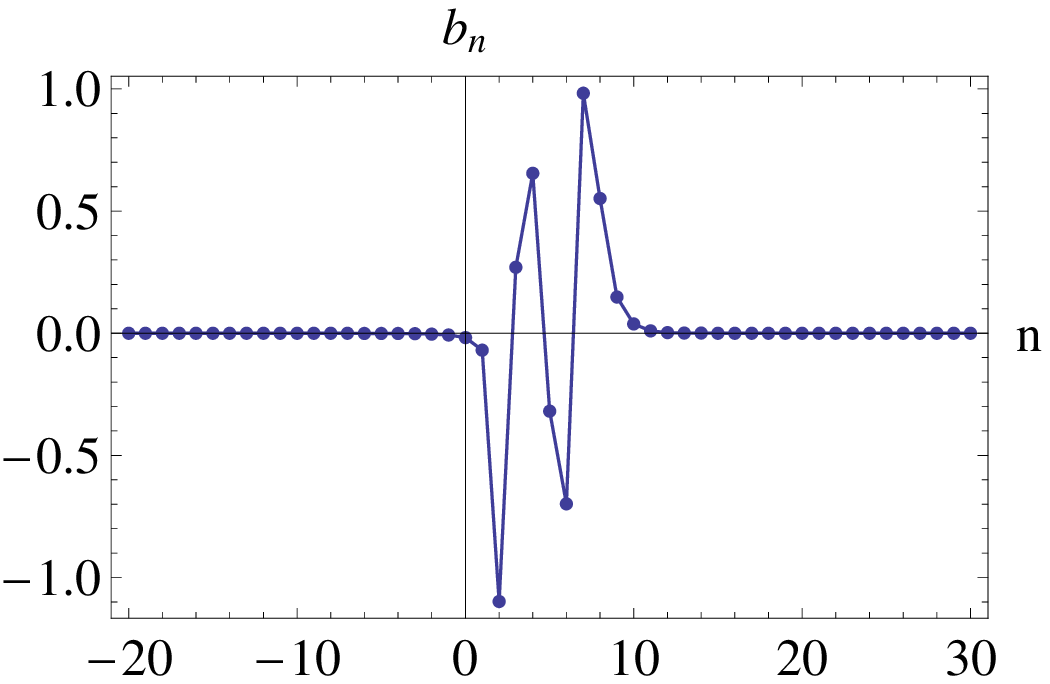}  ~~
\includegraphics[width=3.5cm]{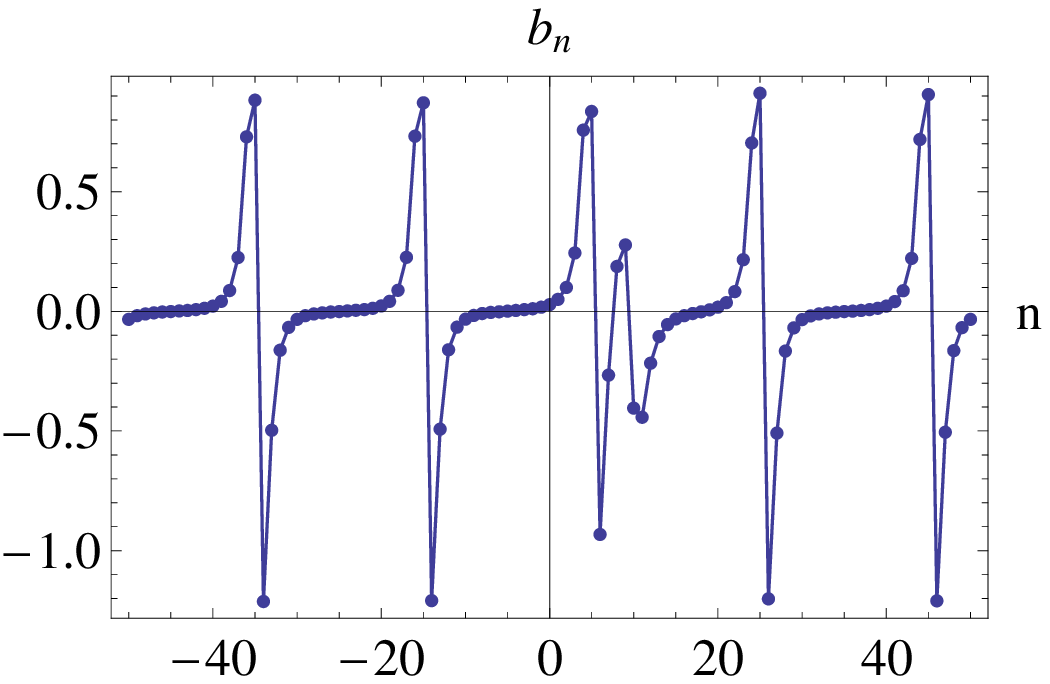} \\
Fig. 4.~~ 2-rational, ~~~soliton-periodic, ~~~soliton-rational,
~~~periodic-rational\\
\end{tabular}
\end{center}

{\bf Example 3}. In this example, we give the positon and negaton
solutions. The term "positon" and "negaton" may be pursued back to
the works of Matveev et al. \cite{MA1}--\cite{MA5}. The positon is
usually given by the trigonometric functions. The positon has some
important properties which differentiates it from the soliton. For
example, it has a square singularity at certain point $x_0(t)$,
and has slow oscillating decay at infinity, and changes its form
with time. The negaton is usually described by the hyperbolic
functions. The nonsingular negaton possesses similar behaviour of
the soliton-like. However, it is different from the soliton. It is
not a translational solution and changes its form with time.
Following the strategy outlined in \cite{MA4}, we can obtain the
positon and negaton solutions for coupled Volterra system (1.5).
We first note that the limit $k_{2}\rightarrow k_1$ leads to the
following Taylor expansion of $\psi^{(1)}_{2,n}$ and
$\psi^{(2)}_{2,n}$:

\begin{subequations}
\begin{equation}
\psi^{(1)}_{2,n}=\psi^{(1)}_{1,n}+\frac{\partial
\psi^{(1)}_{2,n}}{\partial k_2}|_{k_2=k_1}\times(k_2-k_1)+O((k_2-k_1)^2),
\end{equation}
\begin{equation}
\psi^{(2)}_{2,n}=\psi^{(2)}_{1,n}+\frac{\partial
\psi^{(2)}_{2,n}}{\partial k_2}|_{k_2=k_1}\times(k_2-k_1)+O((k_2-k_1)^2).
\end{equation}
\end{subequations}
Then we can rewrite $W^r_n(2)$ and $W^i_n(2)$ as
\begin{subequations}
\begin{equation}
W^r_n(2)=\begin{vmatrix}
 \frac{\partial
\psi^{(1)}_{2,n-1}}{\partial k_2}|_{k_2=k_1}  &  \psi^{(1)}_{1,n-1}  \\
 \frac{\partial \psi^{(1)}_{2,n-2}}{\partial k_2}|_{k_2=k_1} &
\psi^{(1)}_{1,n-2}\end{vmatrix}-\begin{vmatrix}
 \frac{\partial
 \psi^{(2)}_{2,n-1}}{\partial k_2}|_{k_2=k_1}&  \psi^{(2)}_{1,n-1}  \\
 \frac{\partial \psi^{(2)}_{2,n-2}}{\partial k_2}|_{k_2=k_1} &  \psi^{(2)}_{1,n-2} \\
\end{vmatrix}
\end{equation}
\begin{equation}
W^i_n(2)=\begin{vmatrix}
  \frac{\partial
\psi^{(1)}_{2,n-1}}{\partial k_2}|_{k_2=k_1} &  \psi^{(2)}_{1,n-1}  \\
 \frac{\partial \psi^{(1)}_{2,n-2}}{\partial k_2}|_{k_2=k_1}&
\psi^{(2)}_{1,n-2}\end{vmatrix}+\begin{vmatrix}
 \frac{\partial
 \psi^{(2)}_{2,n-1}}{\partial k_2}|_{k_2=k_1}&  \psi^{(1)}_{1,n-1}  \\
 \frac{\partial \psi^{(2)}_{2,n-2}}{\partial k_2}|_{k_2=k_1} &  \psi^{(1)}_{1,n-2} \\
\end{vmatrix}
\end{equation}
\end{subequations}
Taking $\psi_{1,n}^{(1)}$ and $\psi_{1,n}^{(2)}$ as equation (3.6)
or equation (3.5), and using the 2-step Darboux transformation, we
can obtain 1-positon and 1-negaton solutions, respectively.
However, they are too long to write here. We now analyze the
asymptotic behaviour for the positon. We first have
\begin{eqnarray}
W_n^{(r)}(2)&=&e^{t\cos2k_1}[2(c_{12}^{(1)}c_{21}^{(1)}+c_{11}^{(2)}c_{22}^{(2)}-c_{11}^{(1)}c_{22}^{(1)}-c_{12}^{(2)}c_{21}^{(2)})k_1\sin k_1\sin2k_1+(c_{11}^{(1)}\cos\theta+c_{21}^{(1)}\sin\theta)\nonumber\\
&&\times
(c_{22}^{(1)}\cos(\theta-k_1)-c_{12}^{(1)}\sin(\theta-k_1))-(c_{11}^{(2)}\cos\theta+c_{21}^{(2)}\sin\theta)\nonumber\\
&&\times(c_{22}^{(2)}\cos(\theta-k_1)-c_{12}^{(2)}\sin(\theta-k_1))
+(c_{11}^{(2)}c_{12}^{(2)}+c_{21}^{(2)}c_{22}^{(2)}-c_{11}^{(1)}c_{12}^{(1)}-c_{21}^{(1)}c_{22}^{(1)})\eta],
\end{eqnarray}
where $\theta=k_1n-k_1-t\sin2k_1$, $\eta=n-1-2t\cos2k_1$. And
$W_n^{(i)}(2)$ has a completely similar formula. Substituting
$W_n^{(r)}(2)$ and $W_n^{(i)}(2)$ into equation (2.17), and by
careful analysis, we obtain the following asymptotic behaviour of
the 1-positon:
\begin{subequations}
\begin{eqnarray}
a_{n}[2]&=&1-\frac{f_1(\sin\theta,\cos\theta,\sin 2\theta,\cos
2\theta )}{n}, \qquad n \longrightarrow \pm
\infty\\
b_{n}[2]&=&\frac{g_1(\sin\theta,\cos\theta,\sin 2\theta,\cos
2\theta)}{n}, \qquad n \longrightarrow \pm \infty,\\
a_{n}[2]&=&1-\frac{f_2(\sin\theta,\cos\theta, \sin 2\theta,\cos
2\theta)}{t},\qquad  t \longrightarrow \pm
\infty,\\
b_{n}[2]&=&\frac{g_2(\sin\theta,\cos\theta, \sin 2\theta,\cos
2\theta)}{t},\qquad t \longrightarrow \pm \infty,
\end{eqnarray}
\end{subequations}
where  $f_i,$ $g_i (i=1, 2)$ are two rational functions of
variables. The asymptotic behaviour (3.19) yields the conclusion
that the 1-positon of coupled Volterra equation has slow
oscillating decay. We have seen that the singularity structures of
positons are complicated. For example, the positon of the KdV
equation is singular \cite{MA1, MA3}; the positon of the Toda
lattice is 'weakly singular'----- singularity occurs only at some
values of $t$ for every lattice site $n$ \cite{MA4}; non-singular
positons of the non-local KdV equation and the discrete
sinh-Gordon equation have also been found. The analysis of the
singularity structure of the 1-positon presented in this paper is
difficult, since the formula of this positon is very complicated.
By using the method of numerical analysis, for the random values
of $t$ (e.g., $t=-8.45, -2,0, 1.39, 6.78, 20$), we do not find the
singularity of this positon for every lattice site $n$. We thus
think this positon is non-singular. This conclusion is supported
by the Fig. 5 which describes the evolutions of the positon ( with
$c_{11}^{(1)}=c_{21}^{(1)}=c_{11}^{(2)}=1, c_{21}^{(2)}=3;
k_1=2$). We also make the graphs of the negaton (with
$c_{11}^{(1)}=c_{21}^{(1)}=c_{11}^{(2)}=1, c_{21}^{(2)}=-1;
k_1=2$). Obviously, this negaton is not a translational solution
and changes its form with time. N-positon and N-negaton solutions
can be derived through the limit $k_{2i}\rightarrow k_{i}, i=1,2
... N$ and 2N-step Darboux transformations.

Finally, we remark here that for $\alpha>0$ case, setting
$c_{j1}^{(1)}\rightarrow ic_{j1}^{(1)}, j=1,2,$ or
$c_{j1}^{(1)}\rightarrow ic_{j1}^{(1)}, c_{j2}^{(1)}\rightarrow
ic_{j2}^{(1)}, j=1,2$, we can obtain corresponding 1-form and
2-form solutions to coupled Volterra system (1.5).

\begin{center}
\begin{tabular}{ccc}
\includegraphics[width=5cm]{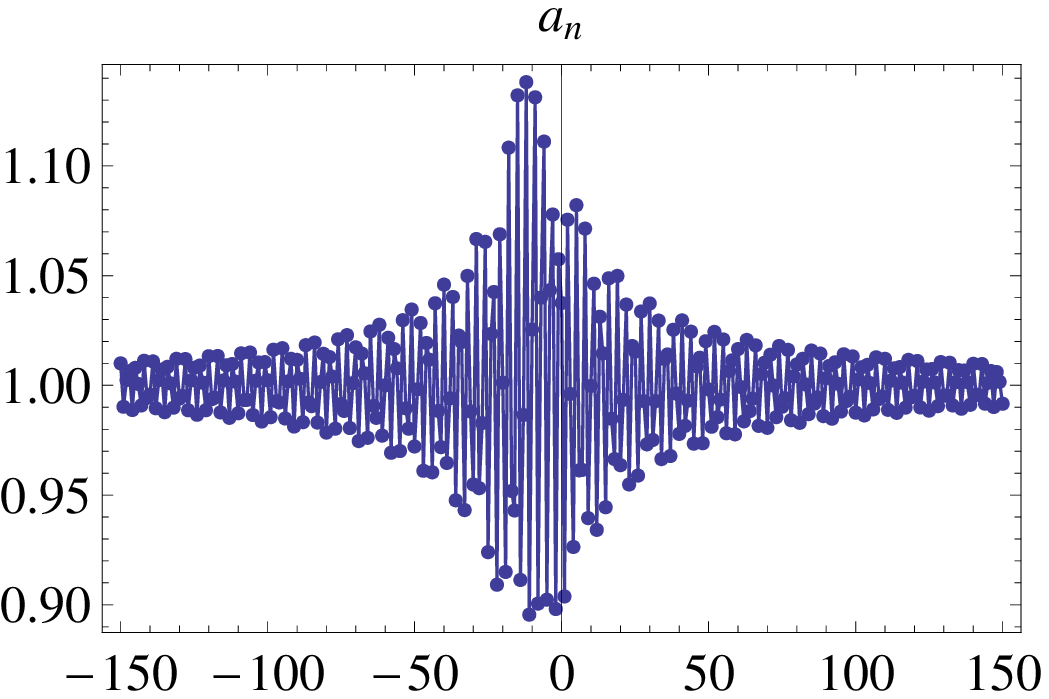} &
\includegraphics[width=5cm]{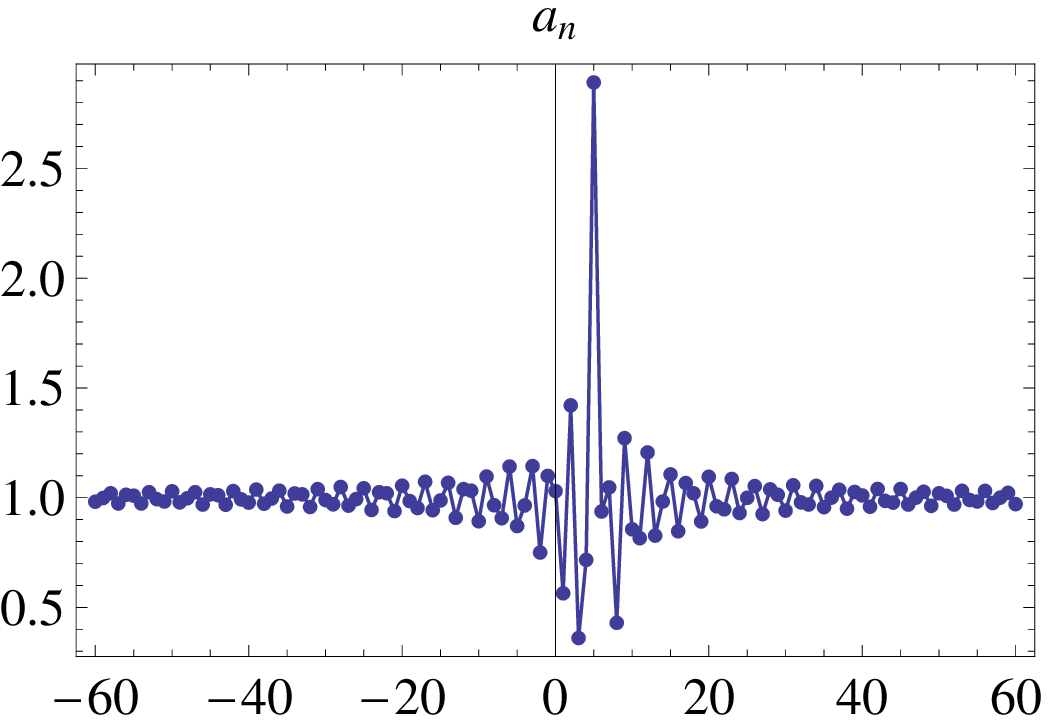} &
\includegraphics[width=5cm]{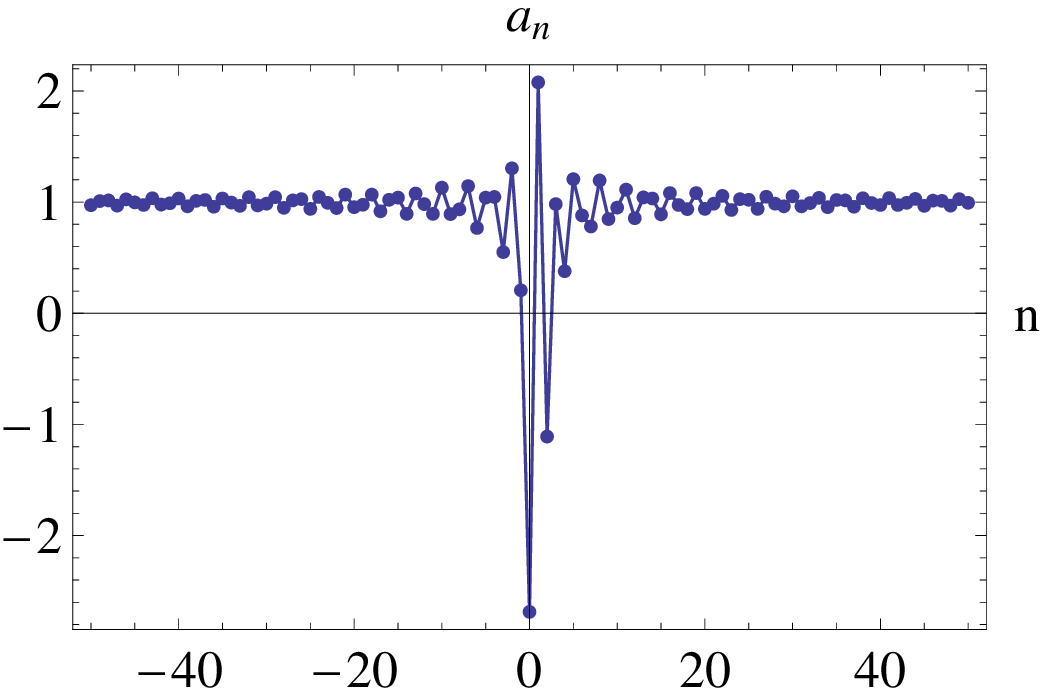} \\
 $a_n(-8.45)$ &  $a_n(-2)$ & $a_n(0)$     \\
\includegraphics[width=5cm]{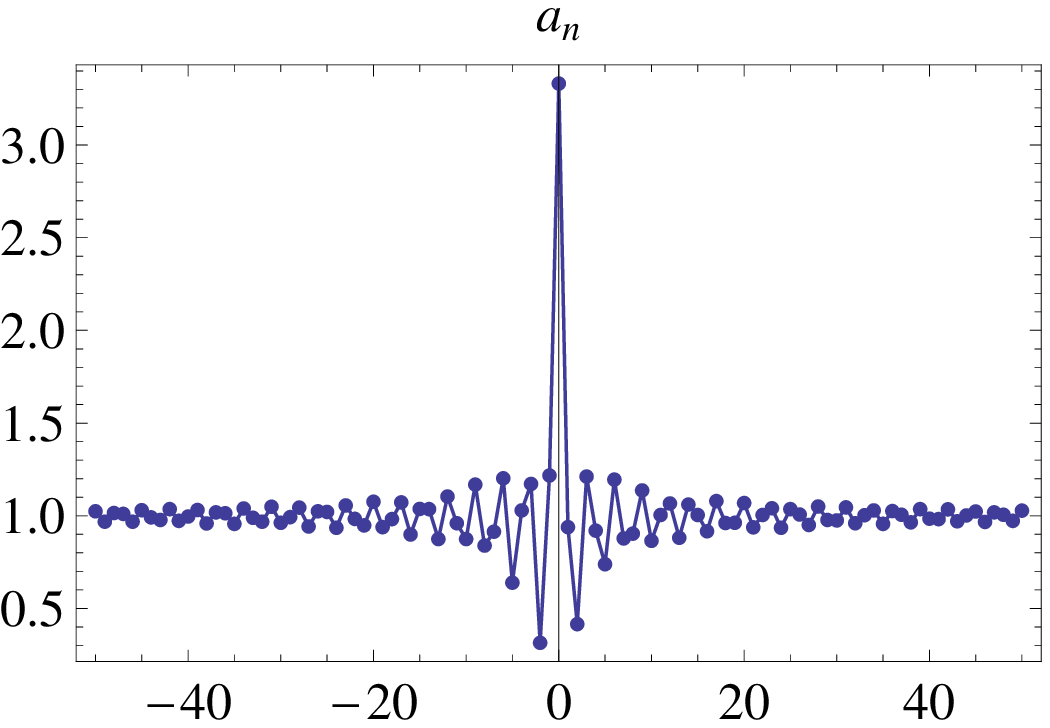} &
\includegraphics[width=5cm]{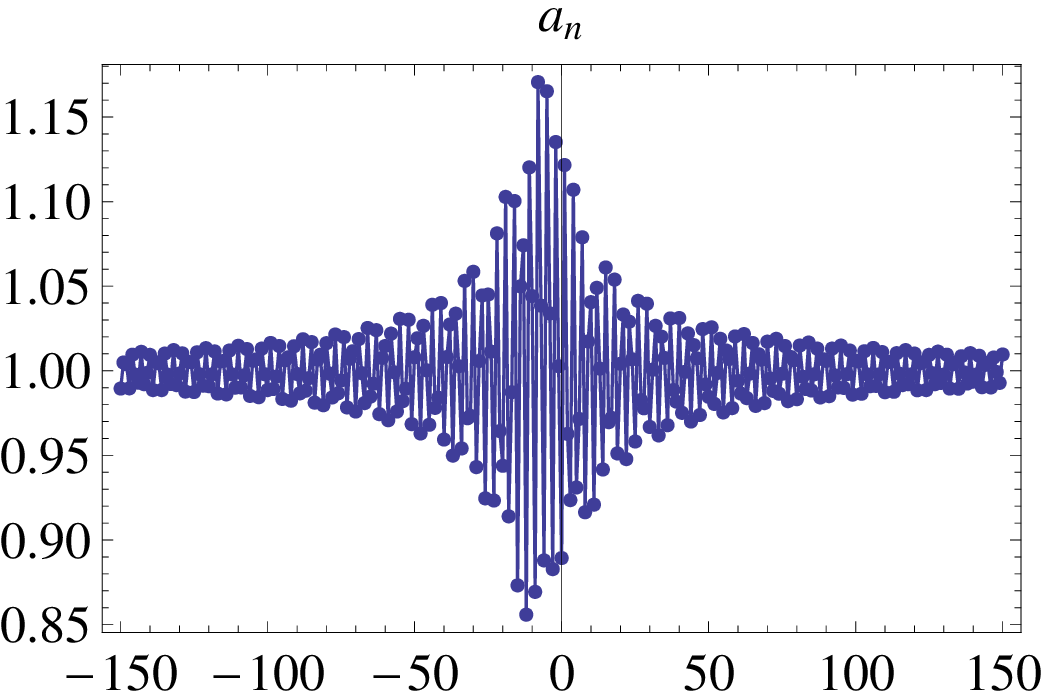} &
\includegraphics[width=5cm]{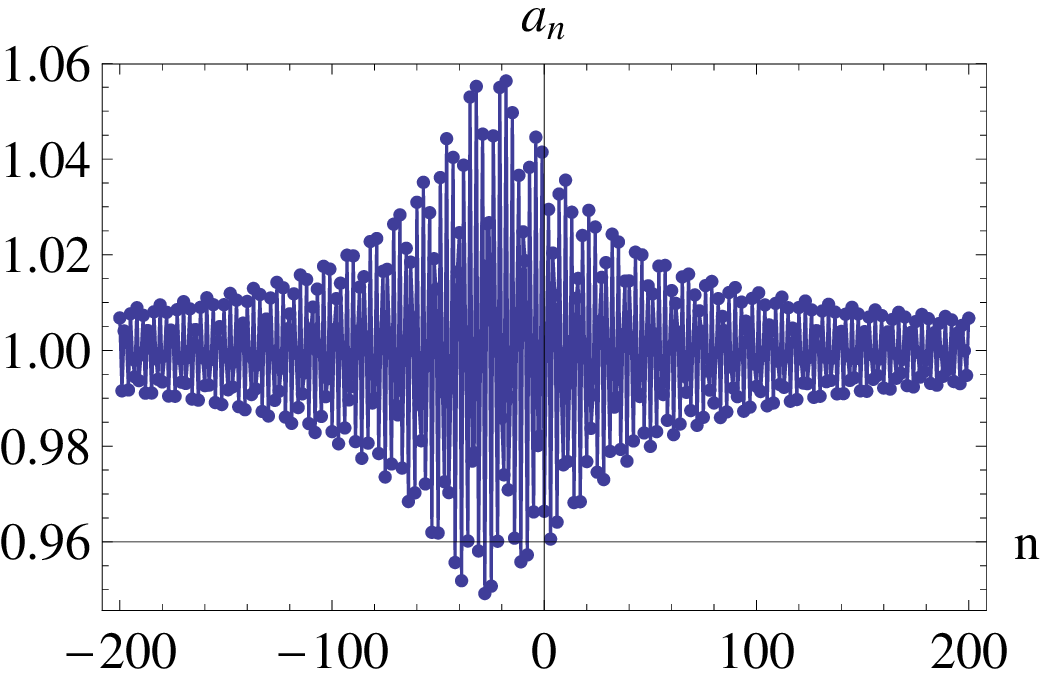} \\
 $a_n(1.39)$ &
$a_n(6.78)$ & $a_n(20)$    \\
\includegraphics[width=5cm]{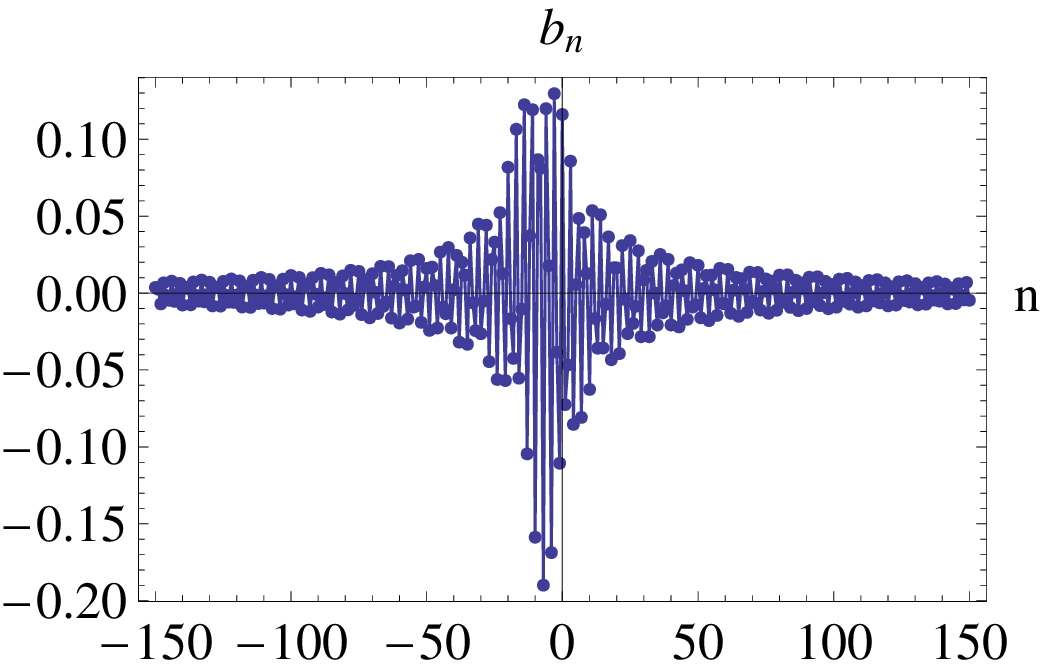} &
\includegraphics[width=5cm]{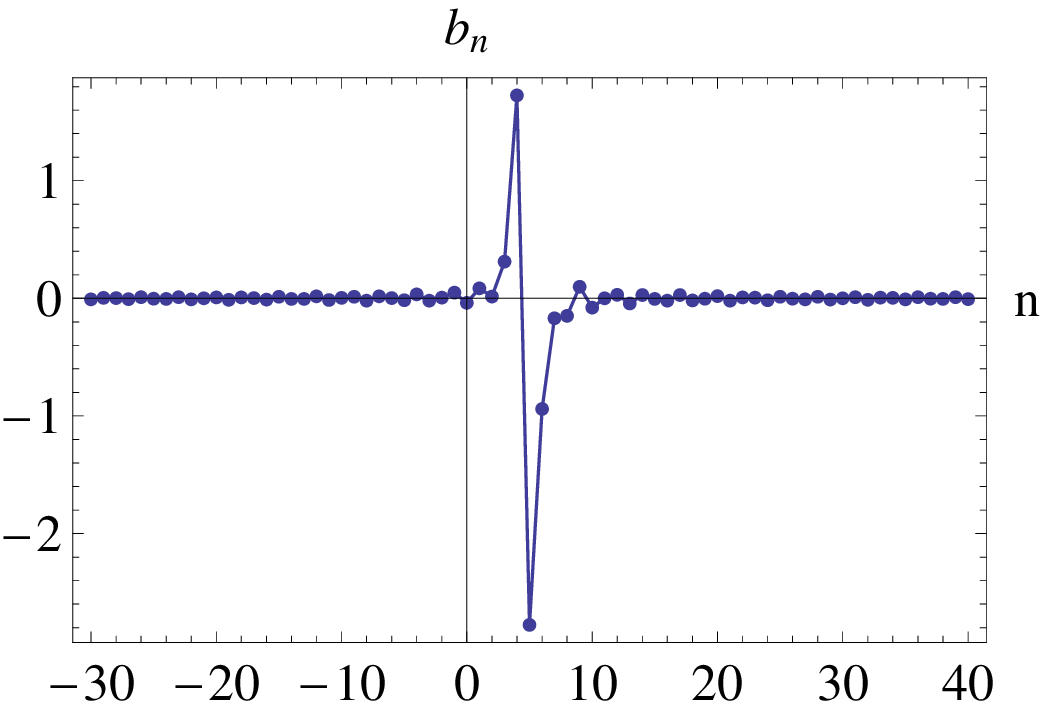} &
\includegraphics[width=5cm]{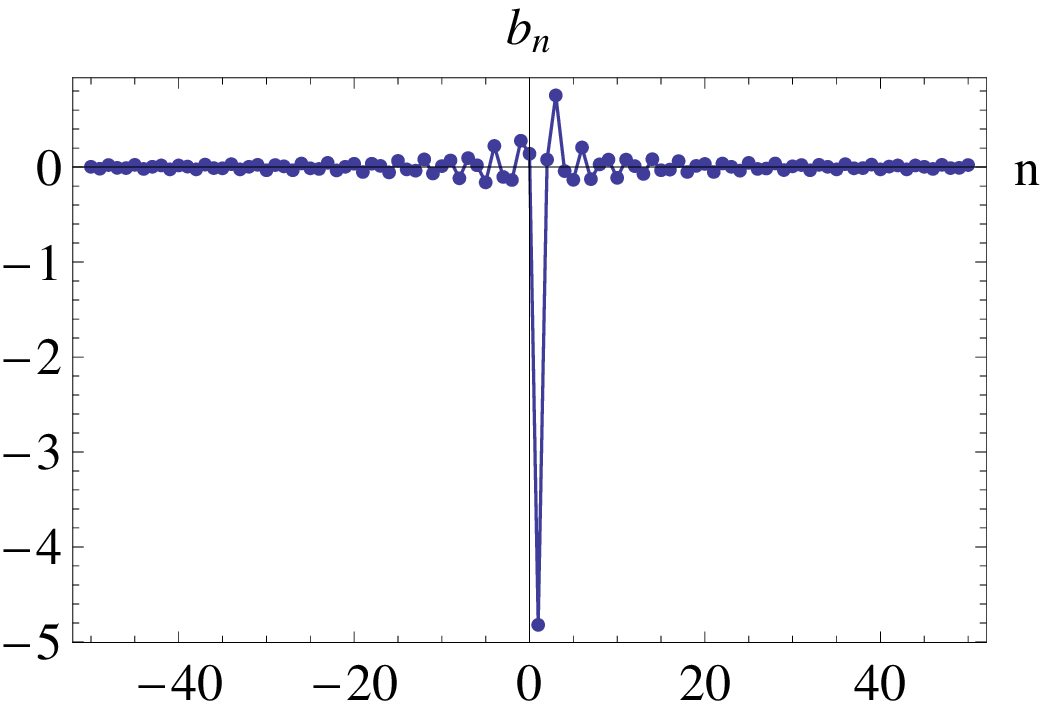} \\
$b_n(-8.45)$   &  $b_n(-2)$ &  $b_n(0)$ \\
\includegraphics[width=5cm]{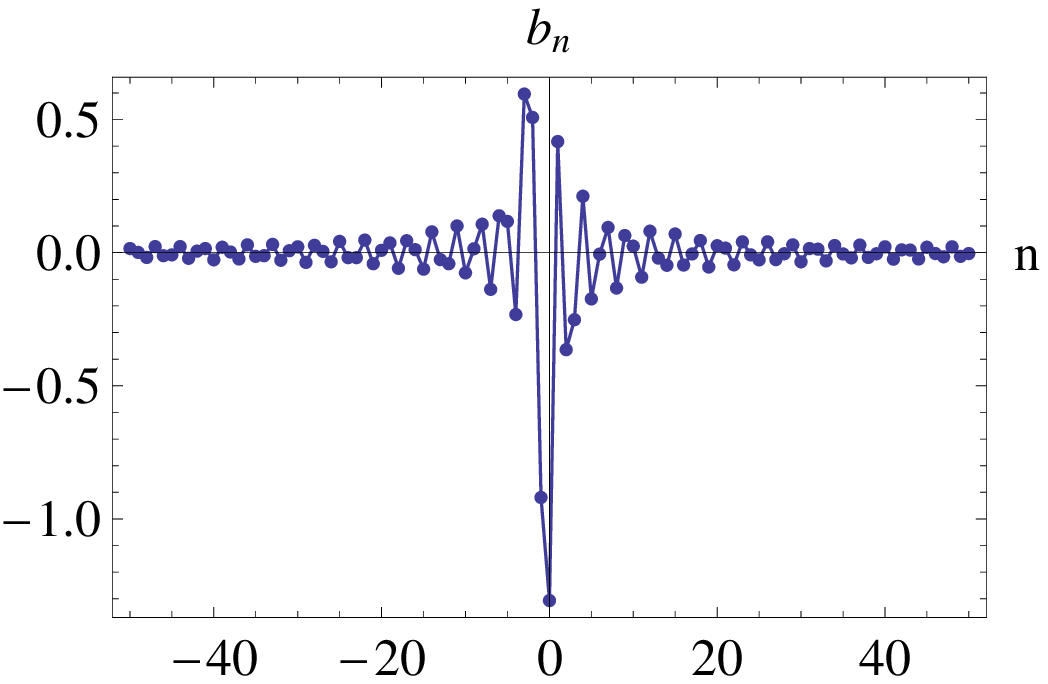} &
\includegraphics[width=5cm]{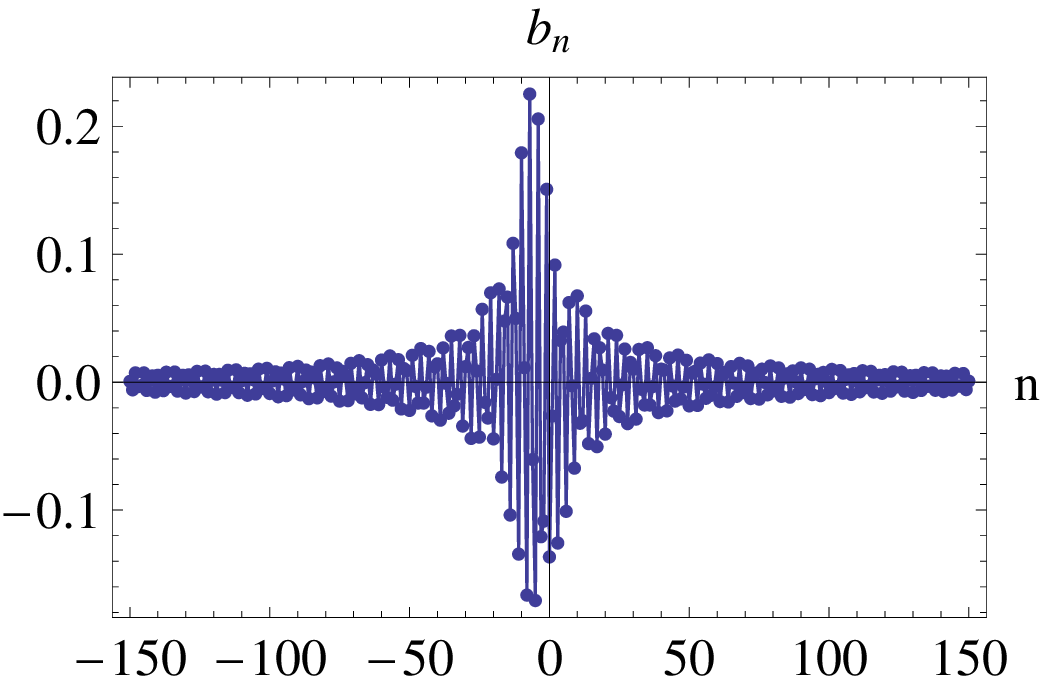} &
\includegraphics[width=5cm]{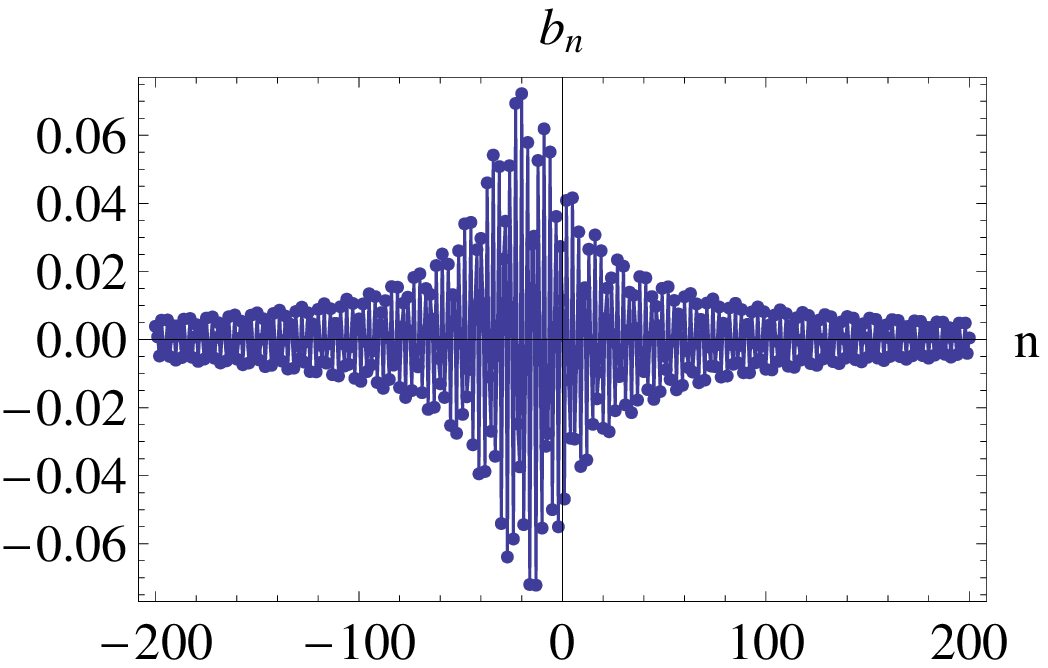} \\
 $b_n(1.39)$ &
$b_n(6.78)$ & $b_n(20)$  \\
& Fig. 5.~~ The evolutions of 1-positon. &
\end{tabular}
\end{center}

\begin{center}
\begin{tabular}{ccc}
\includegraphics[width=5cm]{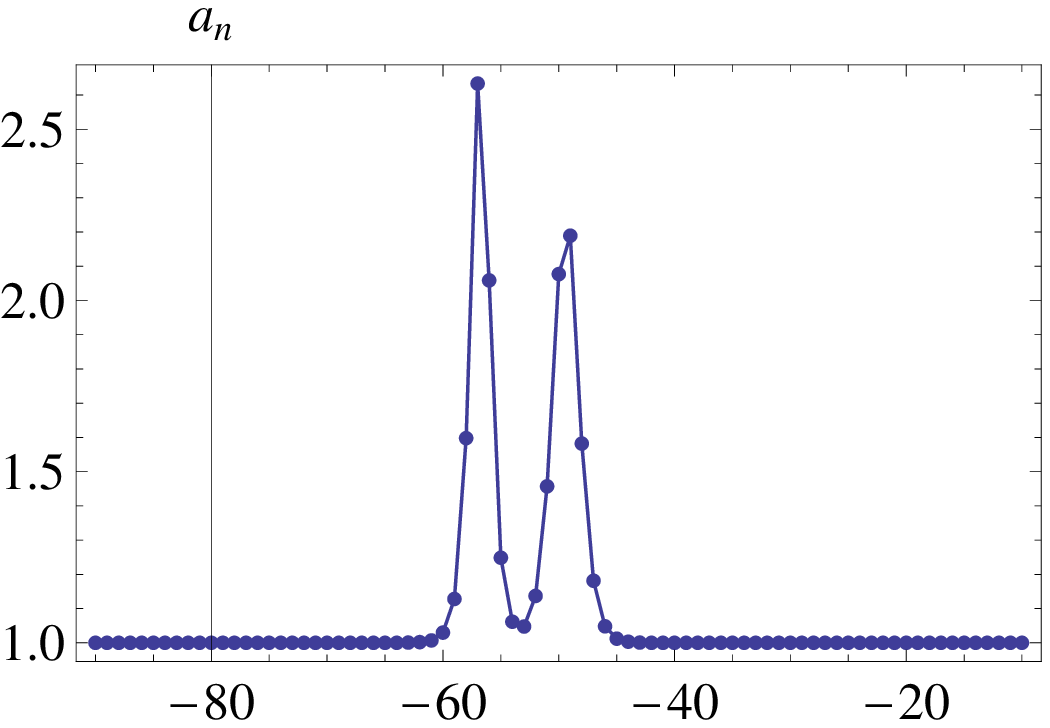} &
\includegraphics[width=5cm]{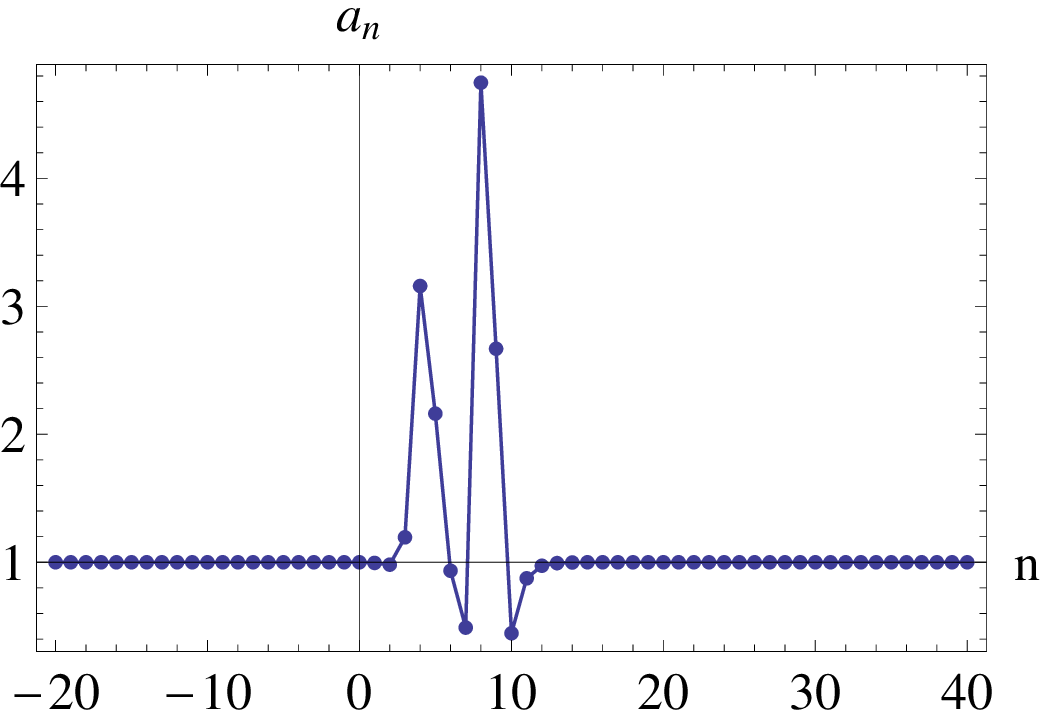} &
\includegraphics[width=5cm]{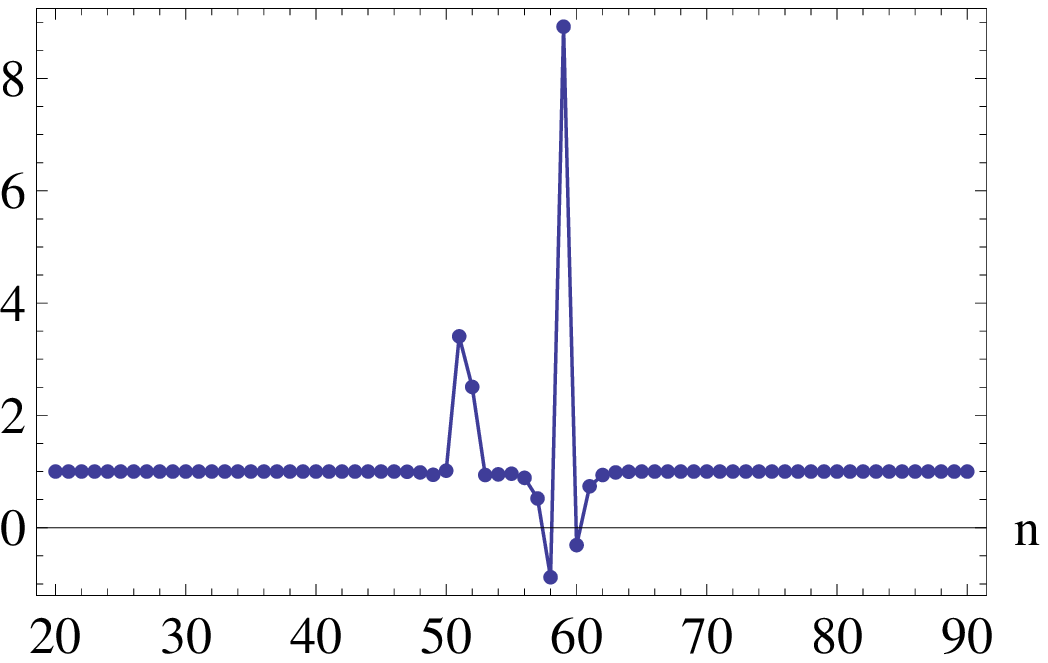} \\
  (a)  &
(b) & (c)  \\
\includegraphics[width=5cm]{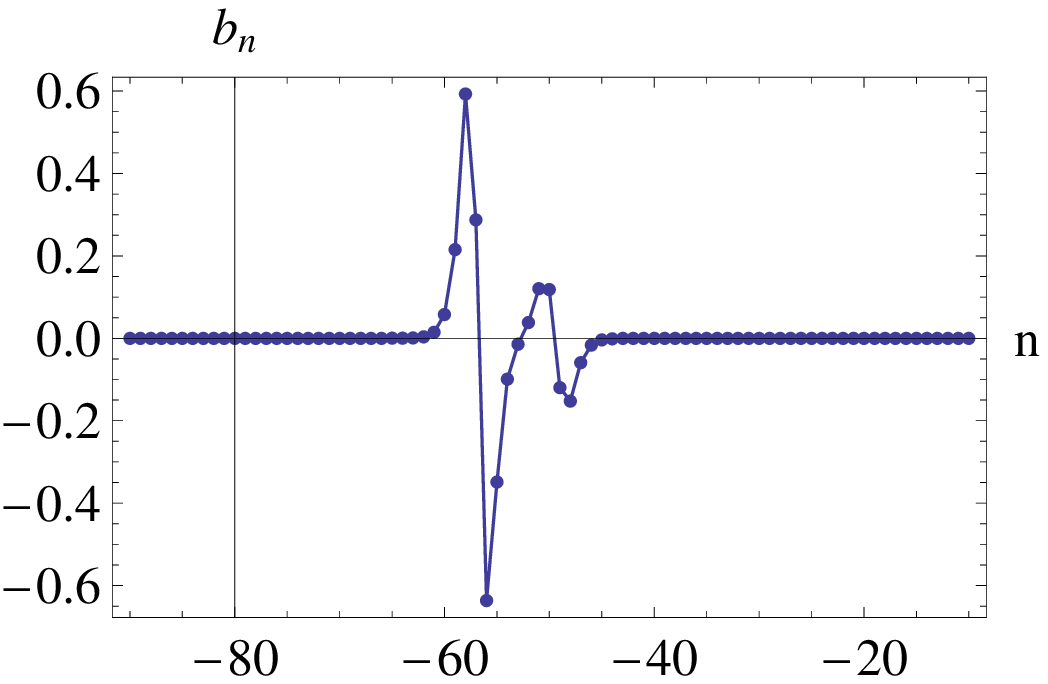} &
\includegraphics[width=5cm]{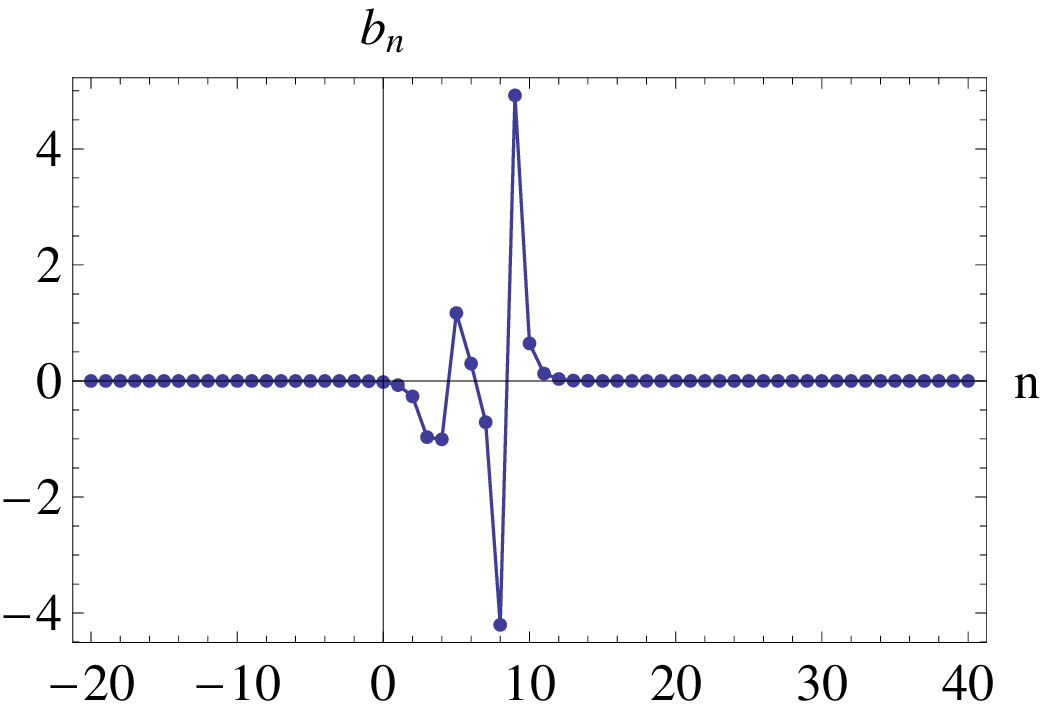} &
\includegraphics[width=5cm]{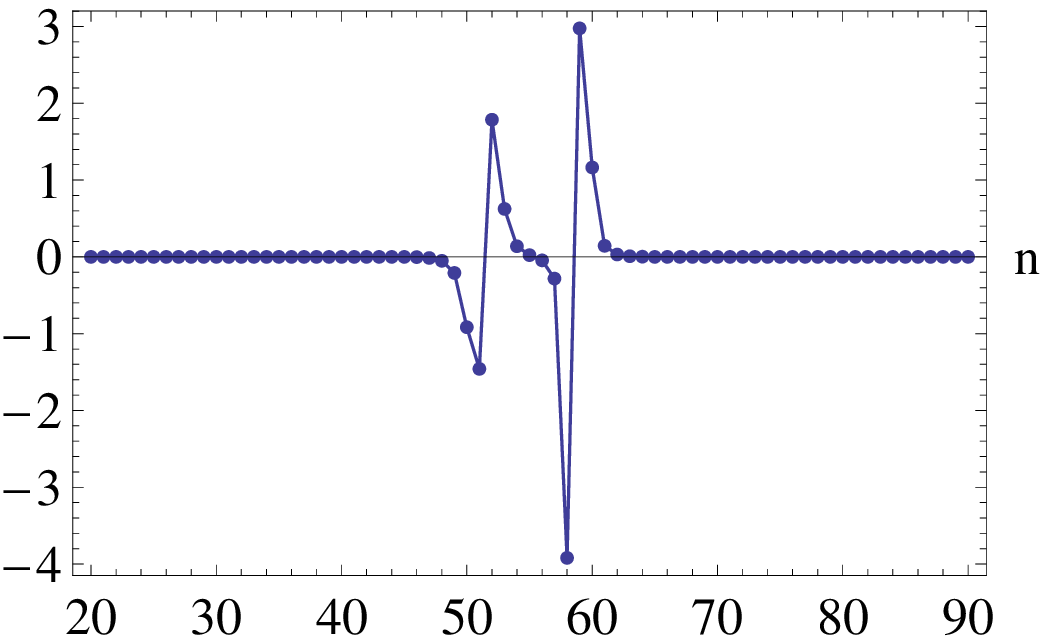} \\
 (d)  &
(e) & (f)   \\
\end{tabular}
Fig. 6.~~ The evolutions of 1-negaton \\ (a),(d) $t=-20$
,~~~(b),(e) $t=2$ ,~~~(c),(f) $t=20$.
\end{center}

\section{Conclusions}

~~~~~~We have found new explicit solutions for a coupled Volterra
system by using the Darboux transformation. These new solutions
include multi-soliton, multi-positon, multi-negaton,
multi-periodic, soliton-periodic, soliton-rational, and
periodic-rational solutions. We have analyzed their dynamical
properties. For example, the multi-periodic solutions are periodic
in both space and time under proper conditions. In the 1-periodic
and 2-periodic cases, the period of the space and time are given.
The asymptotics of the positon is described. The positon obtained
in this paper is a non-singular solution. We also have made plots
for these new solutions. They can help one better understand their
dynamical properties.

\vskip 6mm \noindent {\bf {Acknowledgments}}
 \vskip 1mm The work of
ZNZ is supported by the National Natural Science Foundation of
China under grant 10971136.

\small{

}

\end{document}